\RequirePackage{ifpdf}
\documentclass{JHEP3} 
\pdfoutput=1
\JHEPspecialurl{http://jhep.sissa.it/JOURNAL/JHEP3.tar.gz}
\usepackage{graphicx}
\usepackage{amsmath,amssymb}
\usepackage{cite}
\usepackage{multirow}

\newcommand{\mrm}{\mathrm}
\newcommand{\GeV}{\mathrm{GeV}}
\newcommand{\pT}{p_{\mathrm{T}}}
\newcommand{\pTZ}{p_{\mathrm{T},Z}}
\def\Deltaunc{\Delta_\mathrm{theor.\,unc.}}
\def\Dphill{\Delta \phi_{\ell\ell}}

\def\ConeA{C^Z_{1,\mathrm{A}}}
\def\ConeV{C^Z_{1,\mathrm{V}}}

\def\CtwoA{C^Z_{2,\mathrm{A}}}
\def\CtwoV{C^Z_{2,\mathrm{V}}}
\def\CtwoVA{C^Z_{2,\mathrm{V/A}}}
\def\ConeVA{C^Z_{1,\mathrm{V/A}}}

\def\ttbZ{t\bar{t}Z}
\def\ttbga{t\bar{t}\gamma}

\def\ttb{t\bar{t}}

\def\sw{\sin \theta_w}
\def\cw{\cos \theta_w}
\def\swsq{\sin^2 \theta_w}
\def\ConeVSM{C_\mathrm{V}^{Z,\mrm{SM}}}
\def\ConeASM{C_\mathrm{A}^{Z,\mrm{SM}}}

\def\invfb {\mathrm{fb}^{-1}}

\newcommand{\bal}{\begin{align}}
\newcommand{\eal}{\end{align}}
\newcommand{\be}{\begin{eqnarray}}
\newcommand{\ee}{\end{eqnarray}}
\newcommand{\bspl}{\begin{split}}
\newcommand{\espl}{\end{split}}
\newcommand{\beq}{\begin{equation}}
\newcommand{\eeq}{\end{equation}}

\title{Probing top-Z dipole moments at the LHC and ILC}
\author{Raoul R\"ontsch \\ Fermilab, Batavia, IL 60510, USA \\
  Email: \email{rontsch@fnal.gov} }
\author{Markus Schulze \\ PH Department, TH Unit, CERN, 1211 Geneva 23, Switzerland \\
  Email: \email{markus.schulze@cern.ch} }

\preprint{ CERN-PH-TH-2015-004\\
FERMILAB-PUB-15-010-T}

\abstract{
We investigate the weak electric and magnetic dipole moments of top quark-$Z$ boson interactions at
the Large Hadron Collider (LHC) and the International Linear Collider (ILC).
Their vanishingly small magnitude in the Standard Model makes these couplings ideal for probing 
New Physics interactions and for exploring the role of top quarks in electroweak symmetry breaking.
In our analysis, we consider the production of two top quarks in association with a $Z$ boson at the LHC, 
and top quark pairs mediated by neutral gauge bosons at the ILC.
These processes yield direct sensitivity to top quark-$Z$ boson interactions 
and complement indirect constraints from electroweak precision data.
Our computation is accurate to next-to-leading order in QCD, we include the full decay chain of top quarks and the $Z$ boson,
and account for theoretical uncertainties in our constraints.
We find that LHC experiments will soon be able to probe weak dipole moments for the first time.

}
\begin{document}

\section{Introduction} \label{sec:intro}

Twenty years after the first observation of the top quark, experiments at the Tevatron and the LHC 
have provided us with a detailed picture of top quark properties and interactions.
Today, we have precise knowledge of its basic properties such as the mass, the pair production cross section, spin correlations and the $W$-helicity fractions,
all of which are highly consistent with the Standard Model (SM) predictions.
Moreover, the production of $\ttb$ pairs at the Tevatron and the 7 and 8 TeV LHC spans a wide range of energies
which profoundly probes the QCD production mechanism in different partonic channels, as well as the charged-current decay dynamics.
This has resulted not only in a detailed understanding of the top quark, but has also led to strong exclusion limits for New Physics scenarios.

One aspect of top quark physics which is far less explored are interactions with neutral electroweak gauge bosons and the Higgs boson.
Sensitivity to these couplings arises in hadronic collisions through the partonic subprocess $q\bar{q} \to Z^*/\gamma^* \to \ttb$ 
and through higher-order electroweak corrections to $pp \to \ttb$.
However, at the Tevatron and the LHC the $Z^*/\gamma^*$ mediated process is overwhelmed by the strong production mechanism,
and event numbers at high energies are too low to facilitate a coupling analysis via electroweak loop corrections.
A more promising avenue for anomalous coupling studies is the investigation of the processes $pp \to \ttb+V$ with $V=Z,\gamma,H$
which yield {\it direct} sensitivity without intrinsic dilution by QCD effects or loop suppression.
Their relatively high production threshold of $2m_t+m_V$ requires a sufficiently large center-of-mass energy which 
is, however, adequately reached by the LHC. 
ATLAS and CMS have already measured the $\ttbga$ cross section at Run I \cite{ATLAS:2011nka,CMS:2014wma}, and there is evidence for the production of the $\ttbZ$ final state~\cite{Khachatryan:2014ewa,ATLAS-CONF-2014-038}.  
Run II promises $10^5$ $\ttbZ$ events per 100~$\invfb$ luminosity, and about half as many $\ttb H$ events.
Hence, the study of $\ttb+V$ final states at the LHC will soon open a new chapter in top quark physics
and allow anomalous top quark couplings to electroweak gauge bosons to be probed directly for the first time.
Ultimately these efforts will provide new insights into the role of top quarks in electroweak symmetry breaking
and the effects of New Physics dynamics.\\

Current and past collider experiments already provide some information on anomalous top quark couplings.
For example, electroweak precision data from LEP experiments can be cast into constraints on the parameters
$\varepsilon_1,\,\varepsilon_2,\,\varepsilon_3$ and $\varepsilon_b$ \cite{altarelli:1990zd,Altarelli:1991fk,Altarelli:1993sz}, 
all of which are sensitive to top quark anomalous couplings.
The authors of Ref.~\cite{Eboli:1997kd} use these parameters to constrain the weak magnetic dipole coupling of the top quark.
Flavor changing penguin transitions measured at $B$-factories and kaon experiments are also a sensitive probe of top quark interactions.
The authors of Ref.~\cite{Brod:2014hsa} use rare $B_s$ and $K$ meson decays to constrain vector and axial $\ttbZ$ couplings,
whereas in Refs.~\cite{Kamenik:2011dk,Bouzas:2012av} $B$ decays and asymmetry observables are used to put limits on the electromagnetic dipole moments.
To the best of our knowledge, there are at present no constraints on the weak electric dipole moment.
Although constraints from LEP and heavy meson experiments can be quite strong, it should be noted that they often arise in a convoluted and rather indirect way.
For example, the analysis of Ref.~\cite{Eboli:1997kd} shows that constraints involving the top quark weak magnetic dipole moment
involve three other anomalous parameters, leading to an ambiguous interpretation.
This is not surprising given that top quarks were never observed at LEP and sensitivity only arises through virtual top quarks in electroweak loop corrections.
Similarly, in $Z$ boson mediated penguin transitions the top quarks only enter through virtual exchange and involve additional couplings which can be anomalous.
In contrast, sensitivity in $\ttbZ$ production at the LHC arises directly through interactions of on-shell particles without much dilution by other possible anomalous couplings. 
\\

Looking further into the future, the prospect of an International Lepton Collider (ILC) offers the possibility of extending the LHC top quark program~\cite{Baer:2013cma}. 
The clean collider setting in $e^+ e^-$ collisions and the option for polarized beams presents an optimal 
environment for precision studies of SM processes.
In particular, if the center-of-mass energy is above the top-pair production threshold, the couplings of the top quark 
to the neutral gauge bosons are directly accessible through the process $e^+e^- \to Z^*/\gamma^* \to \ttb$.
 At energies far above the $\ttb$ threshold even the process $e^+e^- \to \ttb+H$ opens up and 
 allows precision studies of the top quark Yukawa coupling.
Moreover, the study of anomalous couplings in $e^+ e^-$ collisions goes beyond the exploration of New Physics models, as it 
is currently also important for estimating the capabilities of a future linear collider. 
Reliable and accurate benchmarks for how well an ILC experiment can pin down fundamental interactions is a major criterion for future planning
that needs to be addressed decades before construction and data taking.\\

In this paper we focus on a particular kind of anomalous couplings: the weak magnetic and electric dipole moments of top quark-$Z$ boson interactions.
In the SM those dipole moments are zero at tree level and  higher-order electroweak corrections introduce values 
which are too small to be measured anytime soon.
Hence the study of anomalous dipole moments provides an ideal place to look for New Physics in the third generation electroweak sector.
Various Beyond the Standard Model scenarios which address the mechanism of electroweak symmetry breaking 
predict sizable dipole moment couplings, as well as deviations of the vector and axial $\ttbZ$ couplings from 
their SM values~\cite{Hollik:1998vz,Agashe:2006wa,Kagan:2009bn,PhysRevD.82.055001,PhysRevD.84.015003,Grojean:2013qca,Richard:2013pwa}.
This raises the question of how well collider experiments can access their parameter space.
This problem has been investigated in Refs.~\cite{Baur:2004uw,Baur:2005wi} for the LHC, and in Refs.~\cite{Devetak:2010na,Amjad:2013tlv} for the ILC.
The former analysis clearly indicates that the leading-order (LO) scale uncertainties are the biggest limiting factor in setting more stringent constraints at the LHC. 
We recently began to address this problem by constraining the vector and axial $\ttbZ$ couplings at the LHC
through next-to-leading order (NLO) in perturbative QCD~\cite{Rontsch:2014cca}, finding 
an improvement by as much as a factor of two compared to a LO analysis.
In this paper we expand our efforts to include the anomalous weak dipole interactions at the LHC, 
and also consider constraints on the $\ttbZ$ couplings at a future ILC with $\sqrt{s}=500$ GeV, both at NLO QCD accuracy.
We also briefly comment on limits achievable in 100~TeV $pp$ collisions. 

The remainder of the paper is organized as follows. In section~\ref{sec:coupl}, we present precise definitions of the $\ttbZ$ couplings, briefly discussing their relationship to operators in an 
effective field theory , and we outline the details of the calculation.
We then consider LHC constraints in section~\ref{sec:LHC} and ILC constraints in section~\ref{sec:ILC}. 
We conclude in section~\ref{sec:concl}.

\section{Anomalous top quark couplings} \label{sec:coupl}
We begin with the Standard Model Lagrangian for $\ttbZ$ interactions
\begin{equation} 
  \label{L_SM}
  \mathcal{L}_{\ttbZ}^{\mrm{SM}} = e \, \bar{u}(p_t)\biggl[ \gamma^{\mu} \bigl( \ConeVSM + \gamma_5 \ConeASM \bigr) \biggr]v(p_{\bar{t}}) Z_{\mu},
  \end{equation}
  where $e$ is the electromagnetic coupling constant. The vector and axial couplings are 
\be
  \ConeVSM = \frac{T^3_t - 2Q_t \swsq}{2\sw \cw} \approx 0.24,  \quad \quad \;
  \ConeASM = \frac{-T^3_t}{2 \sw \cw} \approx -0.60,  
\ee
with $Q_t = 2/3$, $T^3_t=1/2$, and the weak mixing angle $\theta_w$. 
To parameterize possible effects from physics beyond the SM the Lagrangian can be extended to 
\be
  \label{L_NP}
  \mathcal{L}_{\ttbZ} = e \bar{u}(p_t)\biggl[ \gamma^{\mu} \bigl(C^Z_{1,V} + \gamma_5 C^Z_{1,A} \bigr)
  + \frac{\mathrm{i} \sigma^{\mu \nu} q_{\nu}}{M_Z} 
  \bigl(C^Z_{2,V} + \mathrm{i} \gamma_5 C^Z_{2,A} \bigr) \biggr] v(p_{\bar{t}}) Z_{\mu} ,
\ee
with $\sigma^{\mu \nu}=\frac{\mathrm{i}}{2} [ \gamma^{\mu},\gamma^{\nu} ]$ and $q_{\nu} = (p_{t}-p_{\bar{t}})_{\nu}$.
The coupling $C^Z_{2,V}$ corresponds to the weak magnetic dipole moment and $C^Z_{2,A}$ to the CP-violating weak electric dipole moment.
Higher-order corrections within the SM effectively induce finite values for these dipole moments. 
However, their size is very small, in particular $C^Z_{2,V} \approx 10^{-4}$~\cite{Bernabeu:1995gs} and $C^Z_{2,A}$ only receives contributions at 
three-loops~\cite{Czarnecki:1996rx,Hollik:1998vz}.
The anomalous couplings are conveniently parameterized by operators in an effective field theory (EFT) that 
respects the symmetries of the SM. 
Following Ref.~\cite{AguilarSaavedra:2008zc}, the four couplings $C^Z_{1/2,V/A}$ in Eq.~\eqref{L_NP} have a simple translation into four dimension-six operators 
\begin{eqnarray}
%   \label{Cone_NP}
   \ConeV=&C_V^\mathrm{SM}+\left(\frac{v^2}{\Lambda^2} \right) \mathrm{Re} \left[ 2 C^{(3,33)}_{\phi q}  - C^{33}_{\phi u}   \right],\quad 
  & \ConeA=C_A^\mathrm{SM}+\left(\frac{v^2}{\Lambda^2} \right) \mathrm{Re}\left[  2 C^{(3,33)}_{\phi q}  + C^{33}_{\phi u}  \right], \nonumber \\  \\
    \CtwoV =& \sqrt{2} \left(\frac{v^2}{\Lambda^2}\right) \mrm{Re} \left[ c_W C_{uW}^{33}-s_W C_{uB\phi}^{33} \right], \quad \nonumber
  &  \CtwoA = \sqrt{2} \left(\frac{v^2}{\Lambda^2}\right) \mrm{Im} \left[ c_W C_{uW}^{33}-s_W C_{uB\phi}^{33} \right].
  \label{HOOZ}
\end{eqnarray}
 Hence, any constraint on the anomalous couplings can be translated into constraints on a combination of effective operators.
After UV renormalization and evolution of the operators from the EFT scale $\Lambda$ 
down to the electroweak scale, the operators mix at NLO and additional logarithms $\log(M_Z \big/ \Lambda)$ appear.
In this paper we refrain from a comprehensive operator analysis and present constraints on the anomalous couplings only. 
A forthcoming publication on anomalous $\ttbga$ couplings will allow us to come back to this issue in a more general way.

In Ref.~\cite{Rontsch:2014cca} we studied the anomalous couplings $C^Z_{1,\mathrm{V/A}}$ in the process $pp \to \ttbZ$ and treated the $Z$ boson in the narrow-width approximation  
(although spin-correlated decays into leptons were fully included).
In this work, we will investigate the $C^Z_{2,\mathrm{V/A}}$ couplings which involve a momentum-dependent term $\sigma^{\mu \nu} q_{\nu} Z_{\mu}$,
which is to some extent sensitive to off-shell effects of the $Z$ boson.
We therefore extend our previous calculation and allow the $Z$ boson to be off-shell, and include intermediate off-shell photons decaying into leptons.
Consequently, picking up interference between $Z$ and $\gamma$ contributions implies that a possible deviation from the SM couplings can no longer be 
unambiguously traced back to anomalous $\ttbZ$ interactions.
This issue is also apparent in the effective field theory where anomalous $\ttbZ$ and $\ttbga$ interactions share certain effective operators 
(cfg. Eq. (39) and Eq. (41) in Ref.~\cite{AguilarSaavedra:2008zc}) and can lead to correlated shifts.
We will show in the next section that the contribution from intermediate photons in $\ttbZ$ final states at the LHC is negligible as long as the invariant mass 
of the two leptons is not extremely far away from $M_Z$.
Hence, for the scope of this work, we use the SM $\ttbga$ coupling, and treat $\ttbZ$ production at the LHC as a pure and direct probe of top-$Z$ interactions, 
even though intermediate photons are present.
At the ILC, the situation is different.
The intermediate off-shell $Z$ boson and photon mix at the fixed collider energy and thus the contribution from the intermediate photons is not negligible.
Possible avenues to resolving the ambiguity have been explored in Refs.~\cite{Devetak:2010na,Amjad:2013tlv}, facilitating polarized $e^+ e^-$ beams. 
However, in our study of ILC constraints in section~\ref{sec:ILC} we will not address this issue, and will only consider anomalous $\ttbZ$ couplings, 
again keeping the $\ttbga$ couplings at their SM value.
We will, however, come back to this issue in a separate publication on anomalous $pp \to \ttbga$ production at the LHC. \\

On the technical side, our calculation builds upon the results of Ref.~\cite{Rontsch:2014cca}. 
We calculated the NLO QCD corrections to $pp \to \ttbZ$ with leptonic branchings of the $Z$ boson and 
include top quarks decays with  NLO spin correlations.
The top quarks are treated in the narrow-width approximation which is parametric in $\Gamma_t \big/ m_t$
as long as the invariant mass of the top quark is integrated over.
Similar calculations for on-shell tops and parton showered decays have been presented in Refs.~\cite{Lazopoulos:2008de,Kardos:2011na}.
For this study, we extend our  framework by implementing the weak dipole interactions in Eq.~(\ref{L_NP}), 
and by adding a full simulation of the process $e^+ e^- \to t\bar{t} \to  b j j \; \bar{b} \ell^- \bar{\nu} $ 
at NLO QCD, allowing for all possible $\ttbZ$ anomalous couplings.
In this context we briefly mention that the weak dipole couplings in Eq.~(\ref{L_NP}) introduce additional UV-singularities at NLO QCD.
We subtract these divergences through a redefinition of the $C^Z_{2,V/A}$ couplings
\begin{eqnarray} \label{C2counterterm}
    C^Z_{2,V/A} \to C^Z_{2,V/A}\biggl(1 + 
    \frac{\alpha_s}{4\pi} \, C_\mathrm{F} \left(\frac{\mu^2_\mathrm{ren}}{M^2_Z} \right)^\varepsilon \frac{\Gamma(1+\varepsilon)}{\varepsilon}\biggr),
\end{eqnarray}
and confirm their universality in the processes $pp \to \ttbZ$ and $e^+e^- \to t\bar{t}$.
In the EFT this procedure is equivalent to an operator renormalization which arises through the mixing of operators at $\mathcal{O}(\alpha_s)$.

\section{LHC Constraints} \label{sec:LHC}
In this section, we consider the process $pp \to t \bar{t} \;  Z / \gamma^* \to b j j \; \bar{b} \ell^- \bar{\nu} \; \ell^- \ell^+ $
with one $W$ boson decaying hadronically, the other decaying leptonically, and a leptonic branching of the $Z$ boson.
Thus, the final state consists of three charged leptons, four jets (two of which may be $b$-tagged), and missing energy.
In our numerical results we sum over the possible combinations of the charged leptons $\ell^{\pm}=e^{\pm},\mu^{\pm}$.
We concern ourselves exclusively with the LHC operating at a center-of-mass energy $\sqrt{s}=13$ TeV 
and fix the input parameters as follows:
\bal
\label{params}
m_t &= 173~\GeV,    & m_b  &= 0~\GeV, \notag \\
M_Z &= 91.1876~\GeV,& M_W &=80.385~\GeV,\notag \\
G_\mathrm{F} &= 1.166379 \times 10^{-5}~\GeV^{-2}, & \Gamma_Z &= 2.4952~\GeV, \\
\Gamma_t^\mathrm{LO} &= 1.4957~\GeV, & \Gamma_t^\mathrm{NLO} &= 1.3693~\GeV, \notag \\
\Gamma_W^\mathrm{LO} &= 2.0455~\GeV, & \Gamma_W^\mathrm{NLO} &= 2.1145~\GeV. \notag
\end{align}
We use the central factorization and renormalization scale  $\mu_0=m_t + M_Z/2$ and estimate the scale uncertainty by varying it by a factor of two in either direction.
Parton distributions are taken from MSTW2008~\cite{Martin:2009iq} with 
$\alpha_s(M_Z)=0.13939$ and $\alpha_s(M_Z)=0.12018$ at LO and NLO QCD, which we
evolve to the renormalization scale using one-loop and two-loop running, respectively. 
We choose acceptance cuts on the final state to correspond closely to those used in a CMS analysis of the full 8 TeV dataset~\cite{Khachatryan:2014ewa}:
\begin{align}
  \label{selectioncuts}
  \pT^{\ell} & > 20~\GeV,& |y^{\ell}| & < 2.4, \notag\\
  \pT^{j}   & > 30~\GeV, & |y^{j}|   & < 2.4,  \notag \\
  R_{\ell j}&  > 0.3,    & R_{\ell \ell} & > 0.3, \notag \\
  |m_{\ell \ell} -M_Z| &< 20~\GeV. & &
\end{align}
The jets are defined according to the anti-$k_\mathrm{T}$ algorithm~\cite{Cacciari:2008gp} with $R = 0.5$.
With this setup, we obtain a LO and NLO QCD cross-section of 
\begin{align} 
\label{SMcrosssec}
\sigma^{\mrm{LO}} &=2.04^{+36\%}_{-24\%} ~\mrm{fb}, & \sigma^{\mrm{NLO}} &=3.38^{+19\%}_{-18\%}~\mrm{fb},
\end{align}
where the lepton multiplicity of eight has been included. 
The scale uncertainty is obtained by varying the scale by a factor of two in either direction. 
The  $K$-factor of $1.66$ is slightly larger than the one found in Ref.~\cite{Rontsch:2014cca} for the tri-leptonic final state.
This is explained by the more stringent jet cuts used in the present analysis, which are passed more readily when an additional jet is present in NLO kinematics.

\begin{figure}
\includegraphics[scale=0.5]{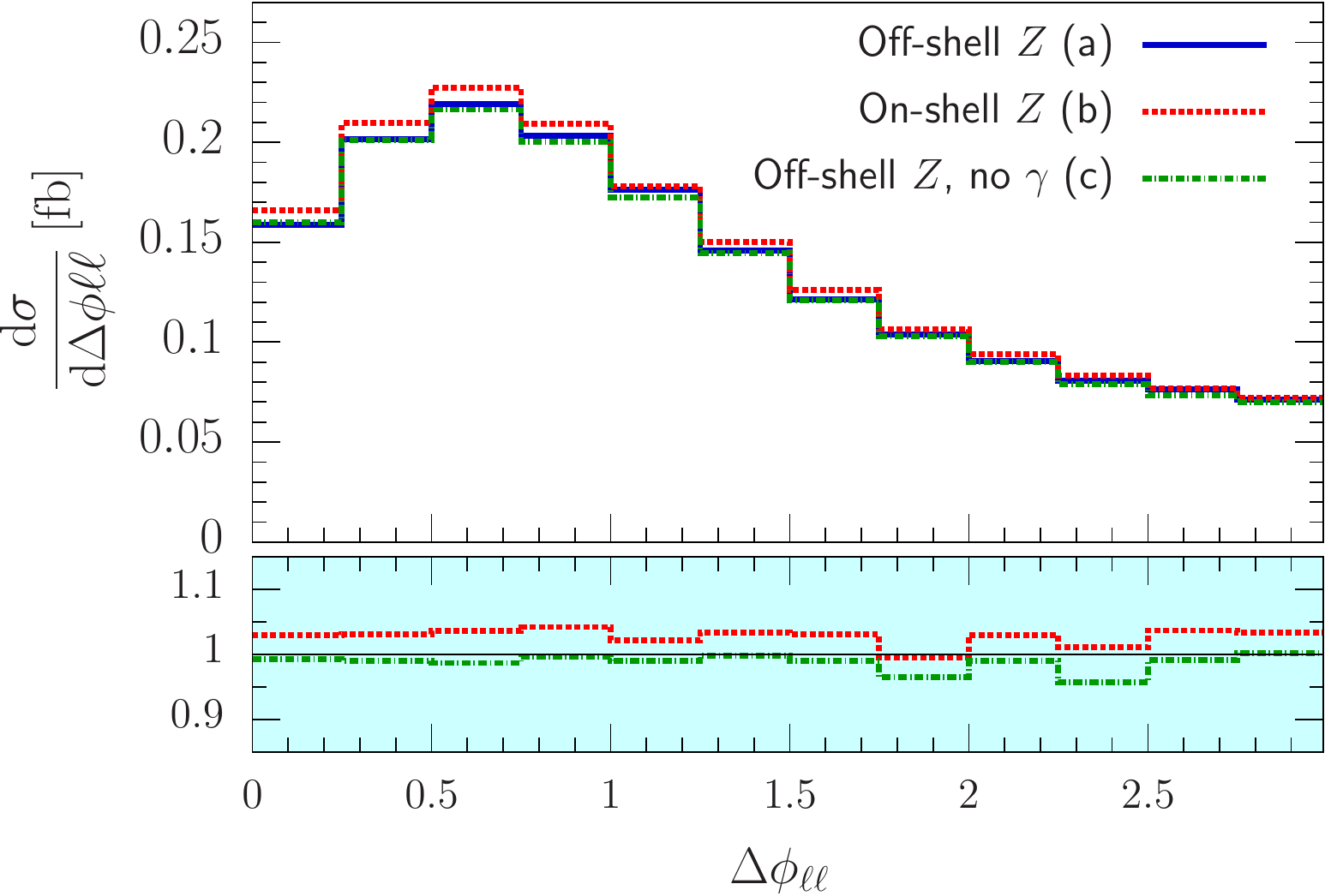}
\includegraphics[scale=0.5]{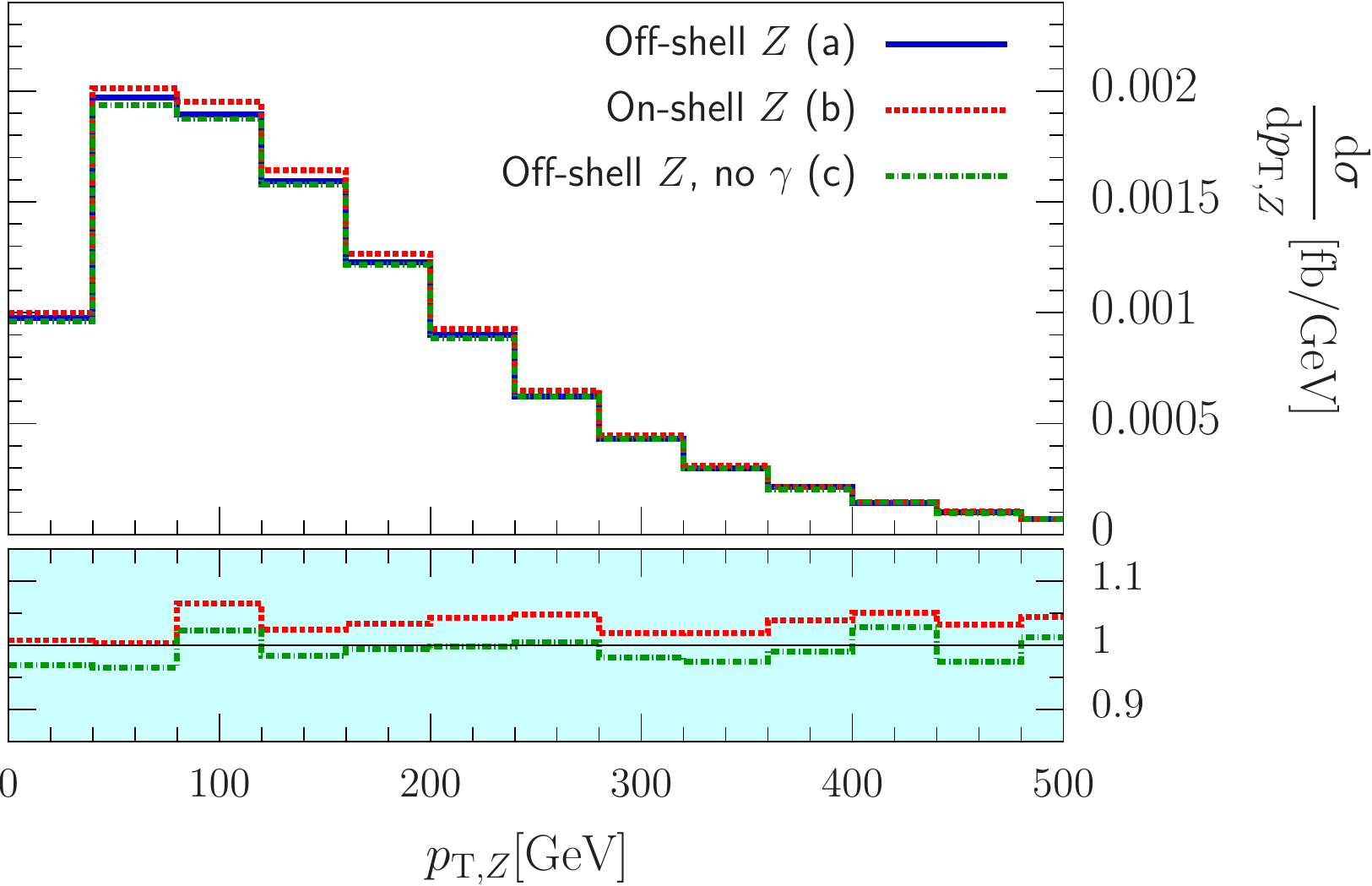}
\caption{NLO QCD distributions of $\Dphill$ and $\pTZ$ for the 13~TeV LHC, using the cuts of Eq.~\protect\eqref{selectioncuts}. 
Shown are the results allowing the $Z$ boson to be off-shell (a), keeping the $Z$ boson on-shell (b), and allowing the $Z$ boson to be off-shell 
but removing the off-shell photon coupling to the top quarks (c). The lower pane shows the ratios (b)/(a) and (c)/(a).\label{fig:offshell}}
\end{figure}
We begin the discussion of results by examining the impact of the $Z$ boson off-shellness and the photon contribution and compare it to the situation in which the $Z$ boson is 
treated in the narrow-width approximation.
In figure~\ref{fig:offshell}, we show two distributions: the azimuthal angle between the two leptons originating from the $Z$ boson decay, $\Dphill$, and the transverse momentum of the $Z$ boson, $\pTZ$.
Allowing the $Z$ boson to go off-shell decreases the cross section by about 3\% at NLO with very little shape change in either distribution.
This is fully consistent with the parametric estimate $\Gamma_Z \big/ M_Z \approx 0.03$ as obtained from the narrow-width approximation.
Removing the contribution of the photon results in a further uniform decrease by 1\%.
This small effect is easily explained by the window cut on the invariant mass of the lepton system, given in Eq.~\eqref{selectioncuts}, which forces the photon to be far off-shell.
This implies that our setup has very little sensitivity to the couplings of the top quark to the photon, 
and we therefore keep $\ttbga$ couplings at their SM values without introducing anomalous interactions.

\begin{figure}
\includegraphics[scale=0.5]{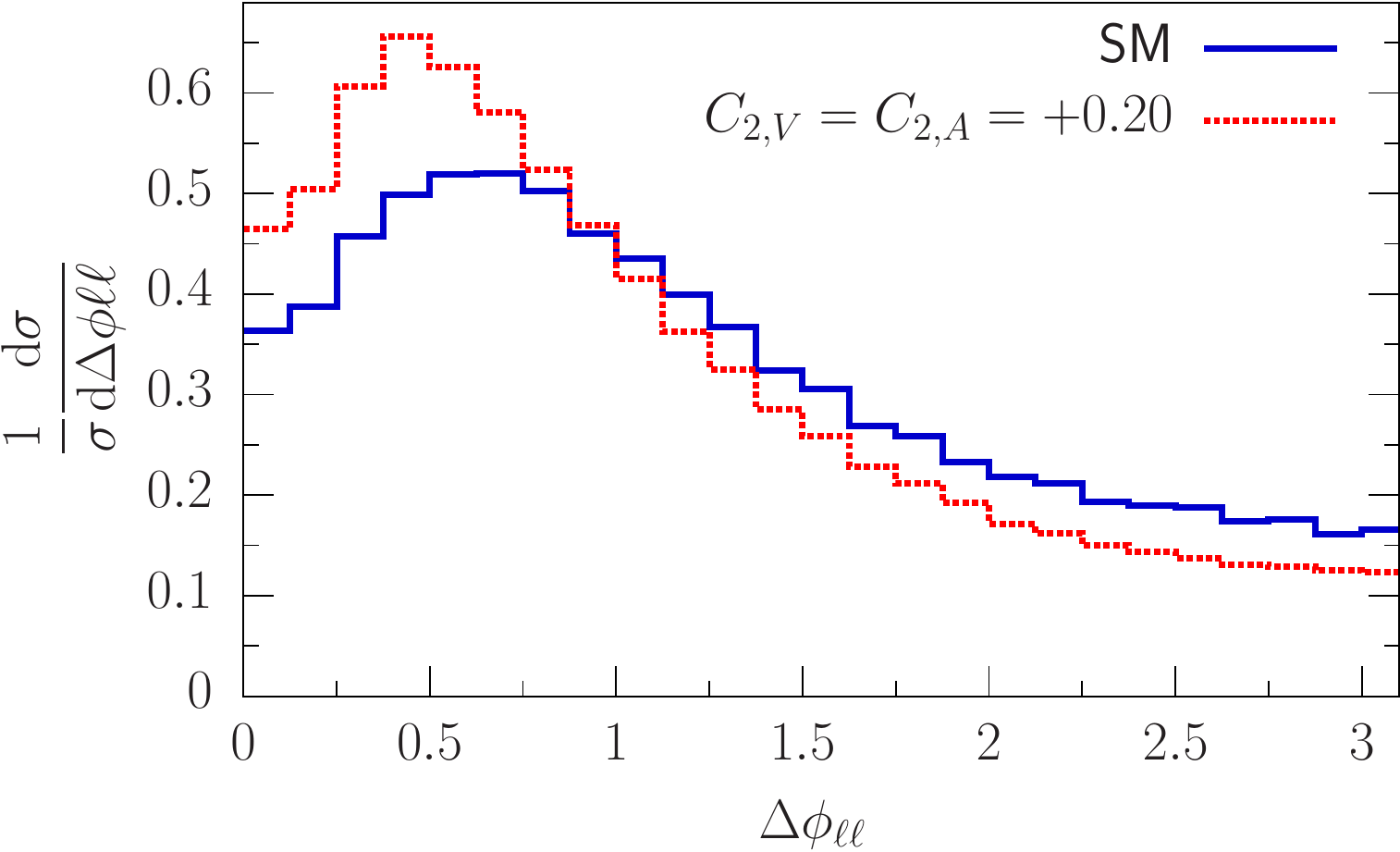} 
\includegraphics[scale=0.5]{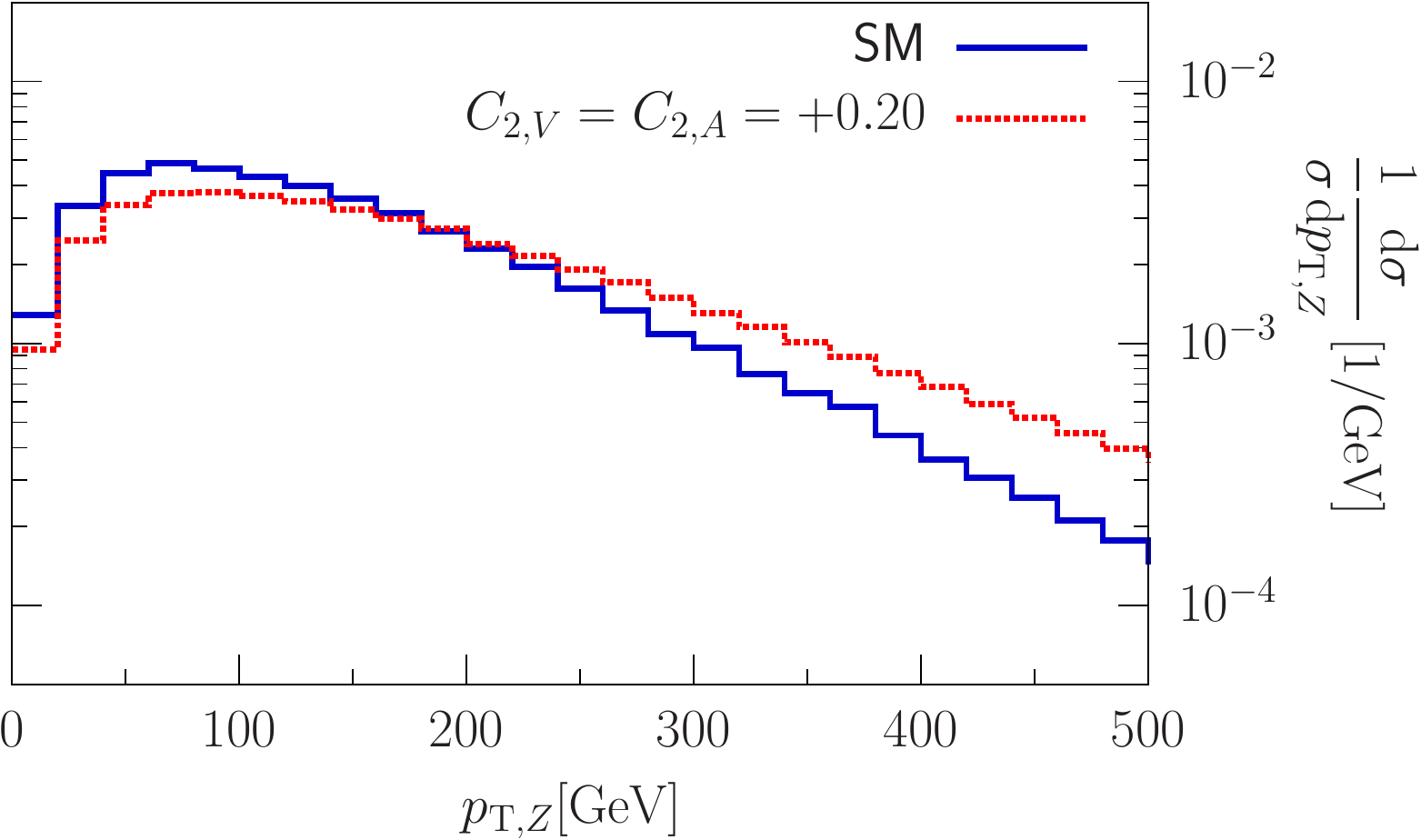} \\
\caption{NLO distributions of $\Dphill$ and $\pTZ$ for SM $\ttbZ$ couplings and with anomalous dipole couplings $\CtwoV=\CtwoA=0.2$. 
The distributions are normalized to the overall cross section. The cuts of Eq.~\protect\eqref{selectioncuts} are applied.} 
\label{fig:LHCdist}
\end{figure}

We now consider the impact of anomalous weak dipole couplings $\CtwoV$ and $\CtwoA$ on the $\ttbZ$ results.
For illustration, we choose $\CtwoV = \CtwoA = 0.2$, but keep $\ConeV$ and $\ConeA$ at their SM values,
and find 
\begin{align}
\sigma^{\mrm{LO}}(\CtwoV=\CtwoA=0.2) &=3.51~\mrm{fb}, & \sigma^{\mrm{NLO}}(\CtwoV=\CtwoA=0.2) &=5.99~\mrm{fb}.
\end{align}
These cross sections are notably larger than the corresponding SM ones in Eq.~\eqref{SMcrosssec}.
Moreover, this choice of anomalous couplings leads to substantial shape changes, as shown in figure~\ref{fig:LHCdist}.
We present the normalized $\Dphill$ and $\pTZ$ distributions and compare anomalous versus SM coupling hypotheses.
The shape changes are significantly stronger than those obtained from anomalous $\ConeV$ and $\ConeA$ couplings as presented in figure~5b of Ref.~\cite{Rontsch:2014cca}.
The reason is clear: varying the values of $\ConeV$ or $\ConeA$ only changes the relative strength of SM vector and axial couplings. 
In contrast, finite contributions of the dipole couplings $\CtwoV$ or $\CtwoA$ introduce a new Lorentz structure to the $\ttbZ$ vertex which 
is proportional to the $Z$ boson momentum vector.
This leads to significantly different spin correlations as well as enhancements in the tail of energy-related distributions.
In the following analysis we will use the two observables presented in figure~\ref{fig:LHCdist} as they provide strong sensitivity to 
the anomalous couplings\footnote{We note that $\Dphill$ and $\pTZ$ are by no means the only observables whose shape has strong discriminating power.
For example, the angle between the semi-leptonically decaying top and the charged lepton originating from this decay in the top quark rest frame is another ideal observable.}. 
The steady enhancement of anomalous cross sections in the $Z$ boson transverse momentum distribution will violate unitarity at very high energies. 
This is related to the transition when the dipole couplings turn into $q^2$-dependent form factors, or equivalently, when the EFT breaks down at energies $\sim \Lambda_\mathrm{EFT}$.
One solution is to introduce a damping factor $(1+q^2/\Lambda^2)^{-2}$~\cite{Baur:2004uw} which prevents unitarity violation at high energies.
In our analysis we simply choose to disregard all normalization and shape information at $\pTZ \ge 500$~GeV.
Assuming that BSM physics enters at scales far above 500~GeV this should be a safe procedure.\\

\begin{figure}[t] 
\includegraphics[scale=0.5]{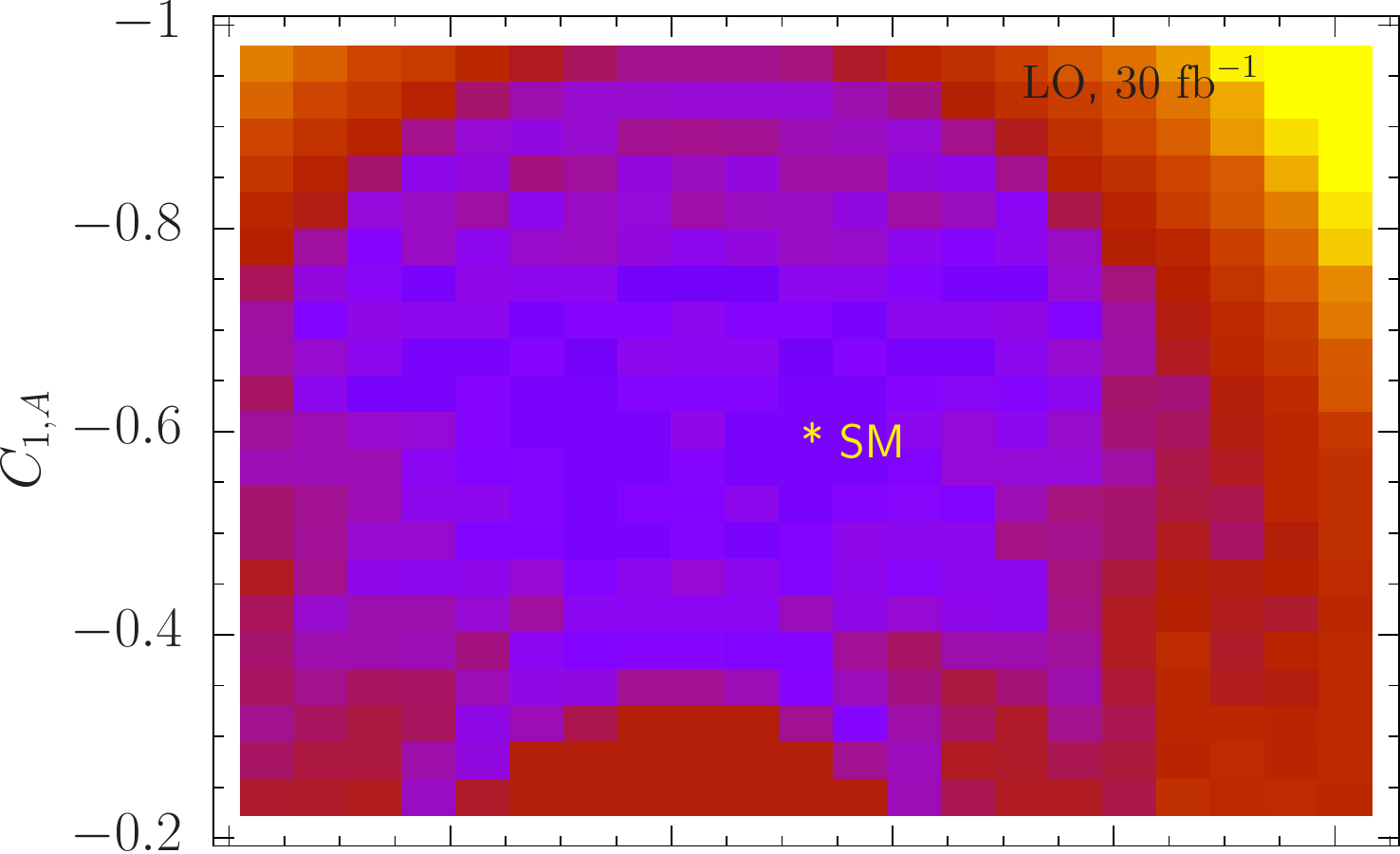}\hspace{0.1cm} 
\includegraphics[scale=0.5]{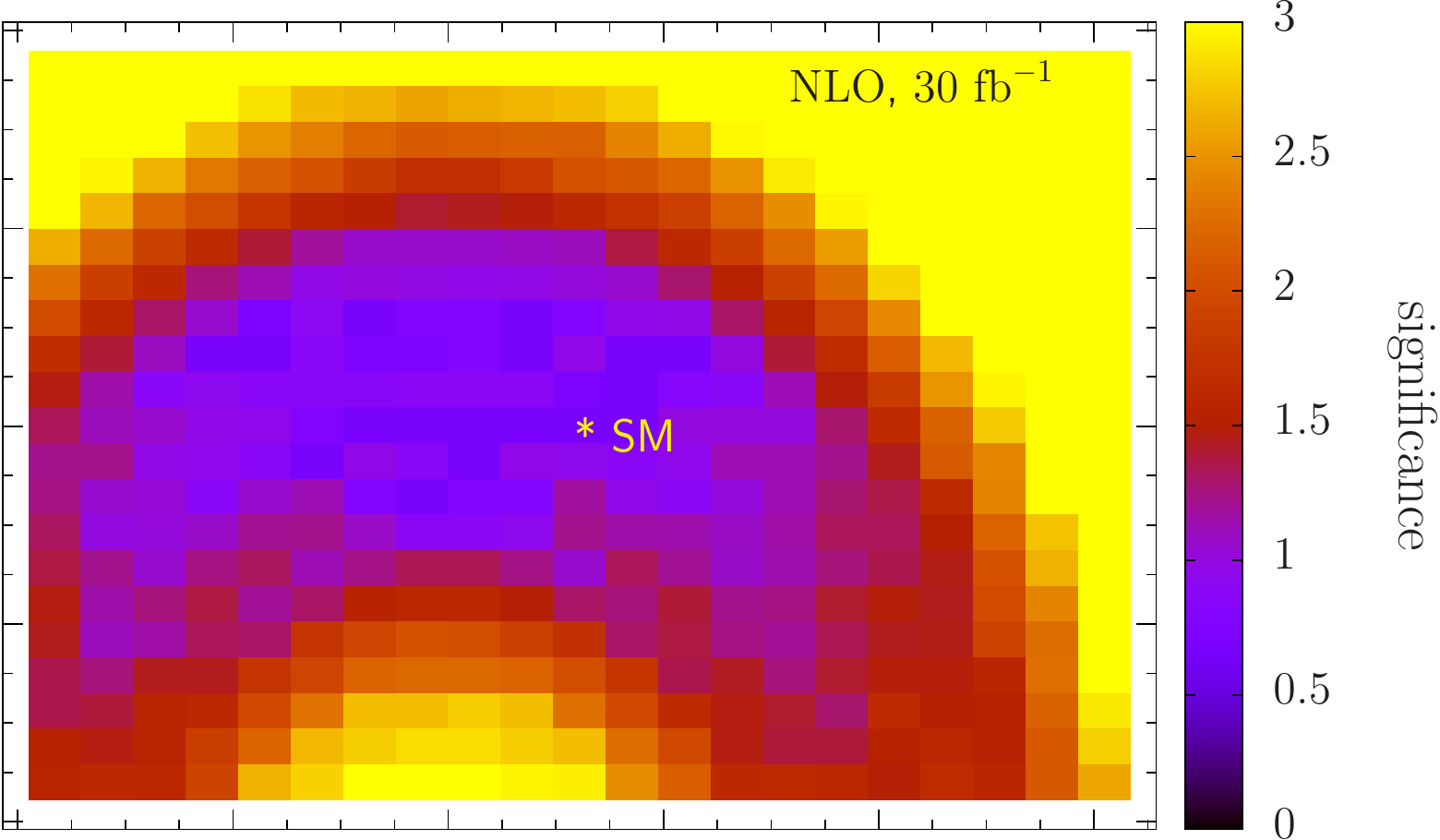}\\
\includegraphics[scale=0.5]{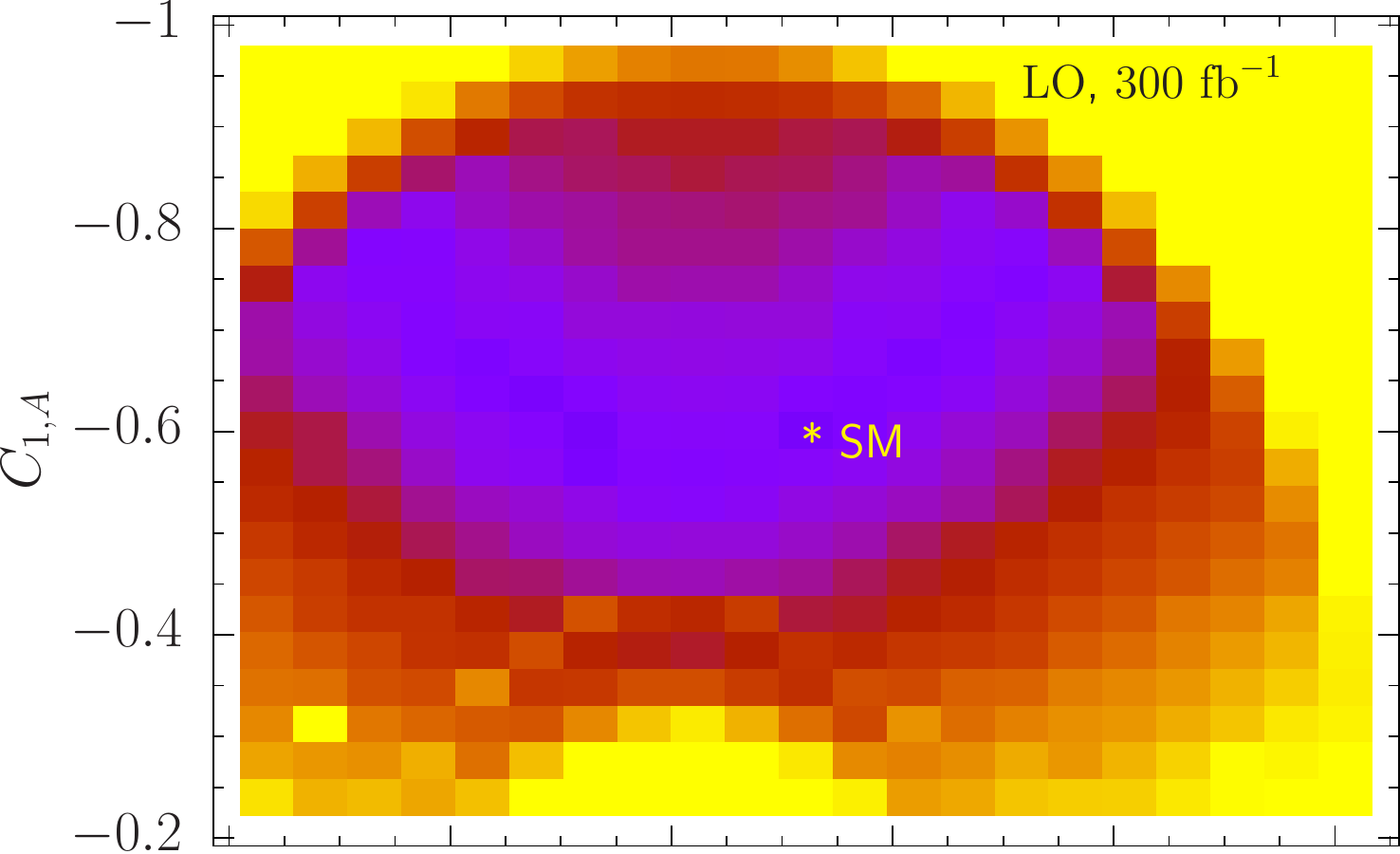}\hspace{0.1cm}
\includegraphics[scale=0.5]{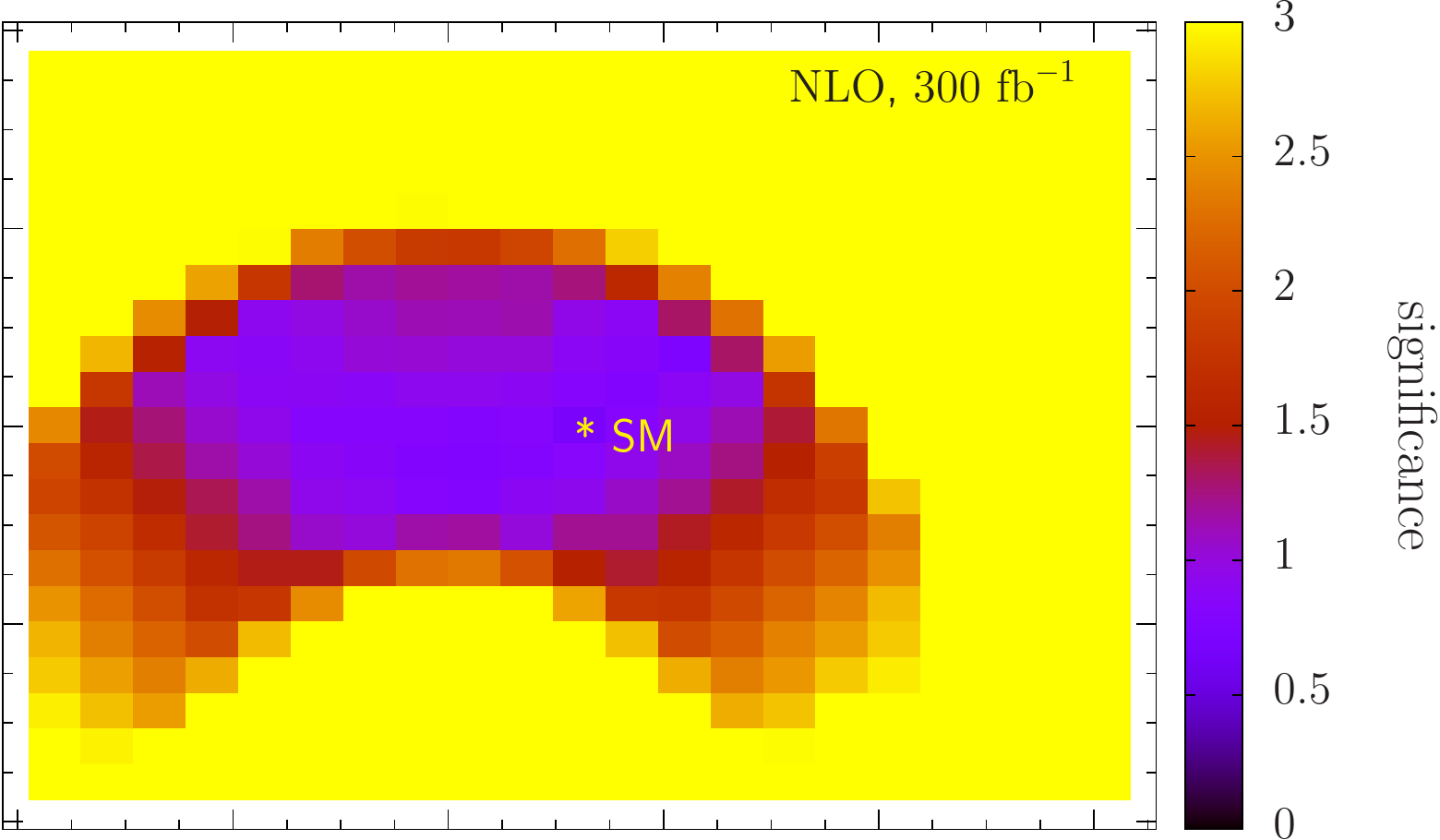}\\
\includegraphics[scale=0.5]{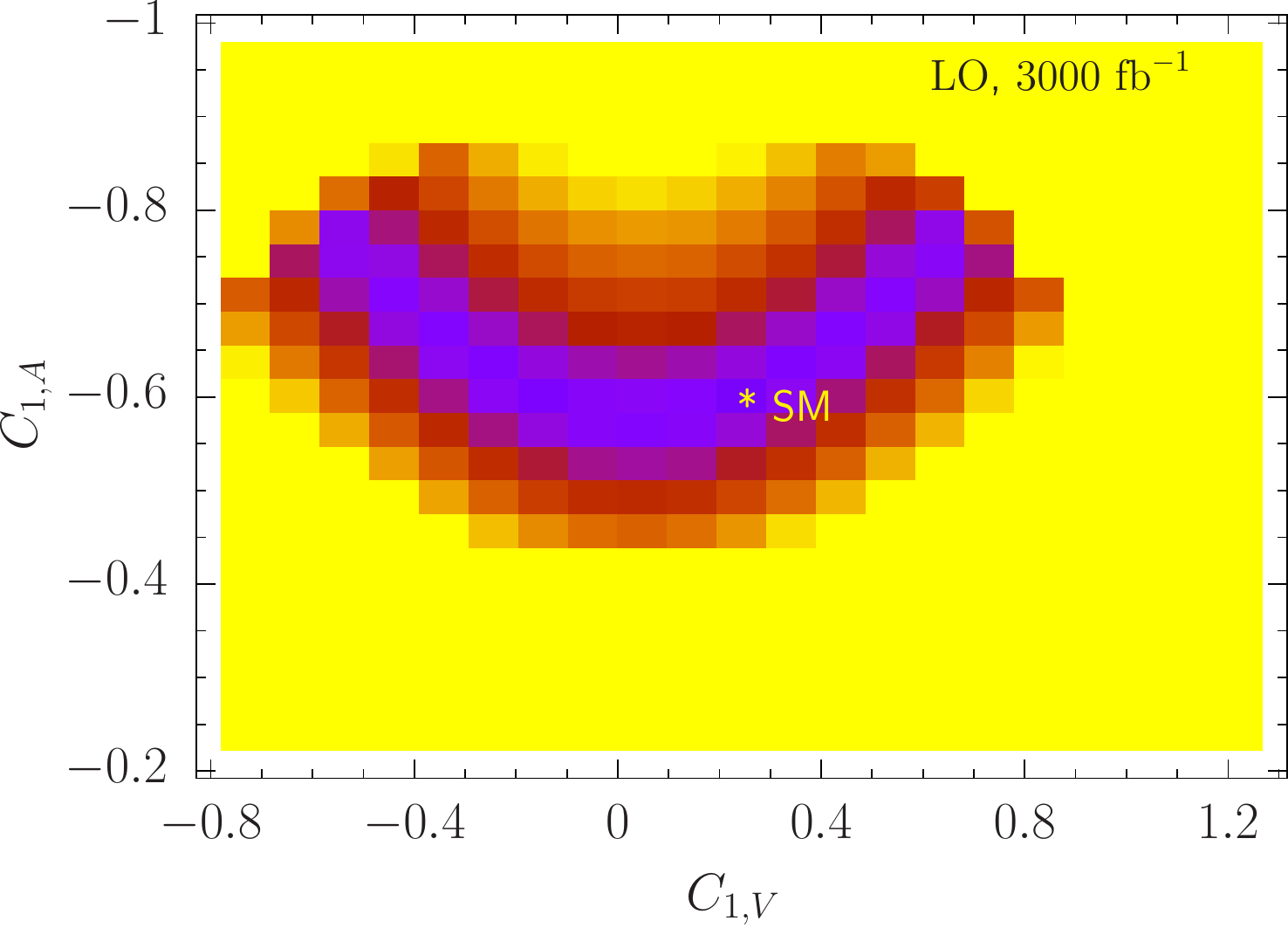}\hspace{-0.05cm}
\includegraphics[scale=0.5]{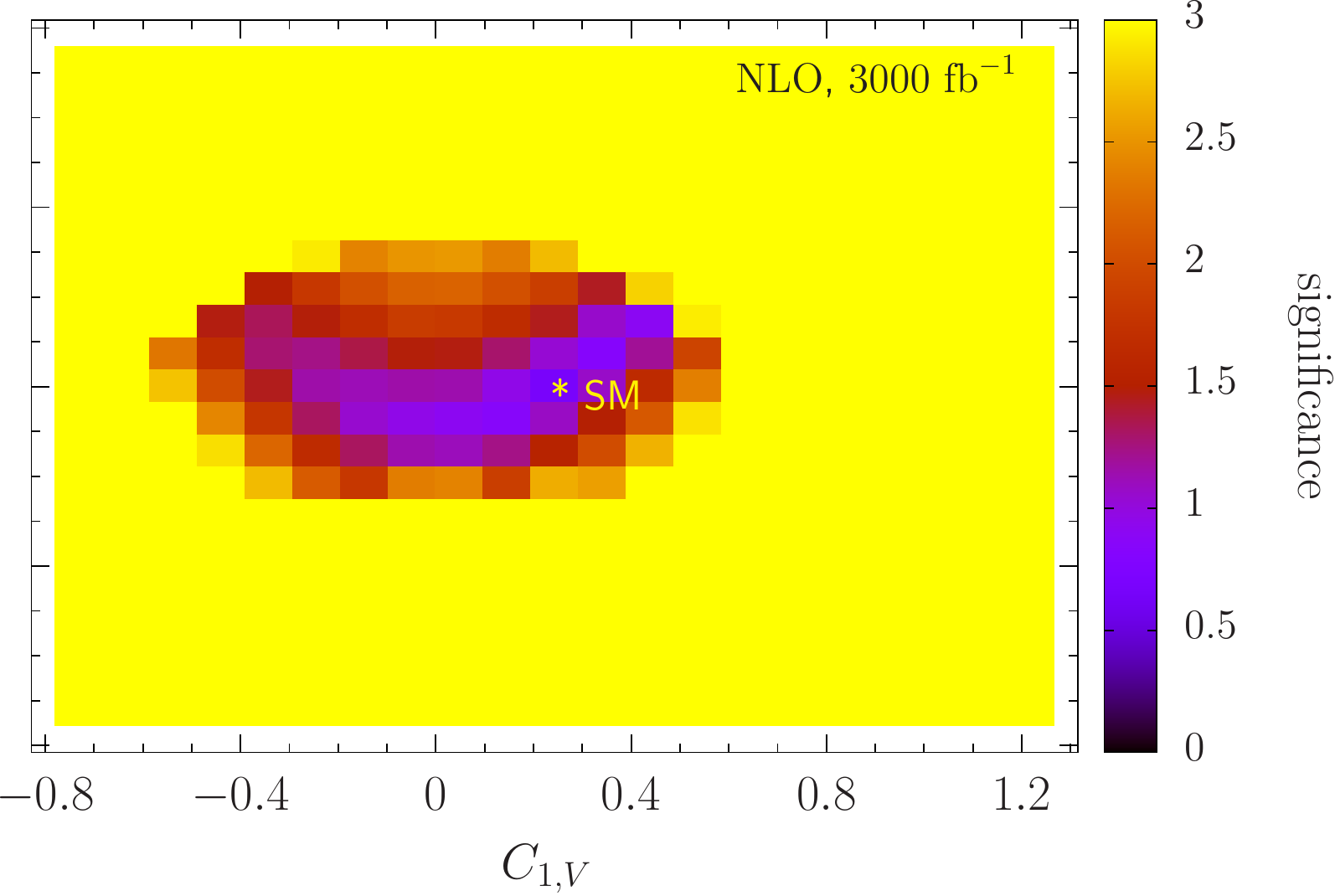}
\caption{Significance of deviations from the SM for the couplings $\ConeV$ and $\ConeA$ with 30, 300, and 3000~$\invfb$ of data at 
the $13$~TeV LHC. The results are obtained from rate and $\Dphill$ shape information using LO predictions shown on the left 
and NLO QCD predictions shown on the right.\label{fig:LHC_C1}}
\end{figure}

We now proceed to studying the constraints that LHC experiments will be able to place on the $\ttbZ$ couplings with 13~TeV data from Run~II.
For reasons of completeness and comparability with the ILC results in section~\ref{sec:ILC}, we first repeat our earlier analysis~\cite{Rontsch:2014cca} by independently varying $\ConeV$ and $\ConeA$ while keeping $\CtwoV$ and $\CtwoA$ zero. 
In fact, we marginally extend the framework of Ref.~\cite{Rontsch:2014cca} by accounting for off-shell $Z$ bosons and photons which were neglected earlier.
%We will independently vary either both $\ConeV$ and $\ConeA$, or both $\CtwoV$ and $\CtwoA$, while the other two couplings are fixed at their SM values. 
We then vary $\CtwoV$ and $\CtwoA$ independently while keeping $\ConeV$ and $\ConeA$ at their SM values.
Thus we will obtain constraints in the parameter spaces of $(\ConeV,\ConeA)$ and $(\CtwoV,\CtwoA)$.
Our strategy closely follows that of Ref.~\cite{Rontsch:2014cca}:
We construct a fitting function to generate LO and NLO kinematic distributions for a large number of anomalous couplings.
We then use a log-likelihood ratio test to determine the significance with which non-SM values of the couplings can be separated from the SM hypothesis, assuming that the SM hypothesis is true.
In the event of the experimental results being compatible with the SM, this can be interpreted as the approximate significance with which a given value of the $\ttbZ$ 
couplings can be excluded.
We include the effects of a theoretical uncertainty $\Deltaunc$ in the likelihood analysis through the overall rescaling of the results under both hypotheses, such that the difference between the two cross sections is minimized. 
The value of the  theoretical uncertainty is taken to be $\Deltaunc= 30\%$ at LO and $\Deltaunc = 15\%$ at NLO, corresponding roughly to the scale uncertainties from missing higher order corrections.

The constraints on $\ConeV$ and $\ConeA$, obtained using the $\Dphill$ distribution, are shown in figure~\ref{fig:LHC_C1} for an integrated luminosity of 30, 300, and 3000~$\invfb$.
The area outside of the blue area roughly corresponds to couplings that can be excluded at 68\% confidence level (C.L.); the area outside the red area can be excluded at 95\% C.L.
Thus we expect coupling constraints, using 300~$\invfb$ of data and employing NLO QCD results, of approximately 
$-0.75 \lesssim \ConeA\!=\!-0.60 \lesssim -0.5$ and $-0.7 \lesssim \ConeV\!=\!0.24 \lesssim 0.7$ at the 95\% C.L.
The improvement when using NLO QCD over LO predictions is immediately apparent from the figures. 
This is mainly a consequence of the reduced scale uncertainty and the larger event rate due to a positive perturbative correction.
\\
\begin{figure}
\includegraphics[scale=0.5]{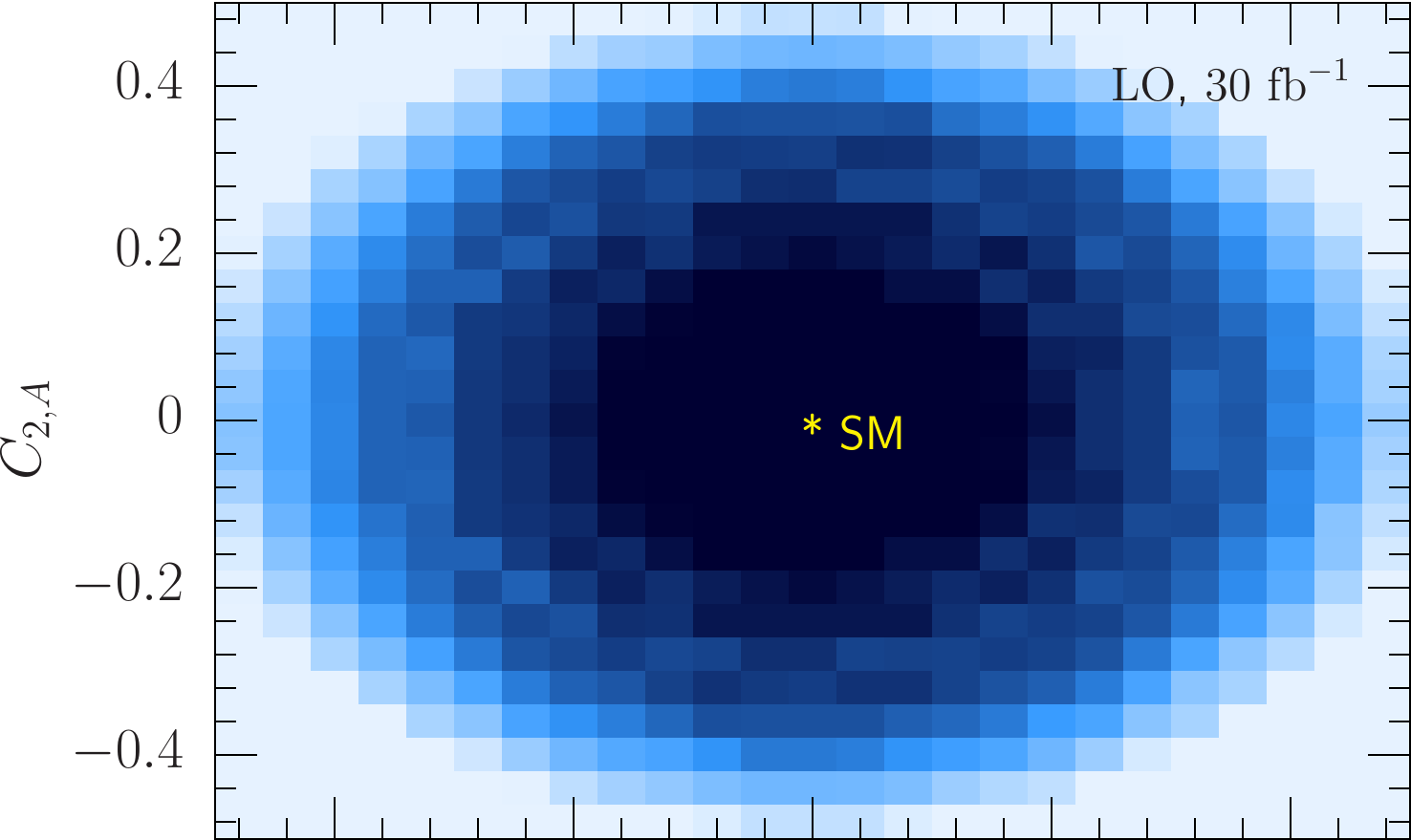} \hspace{0.1cm}
\includegraphics[scale=0.5]{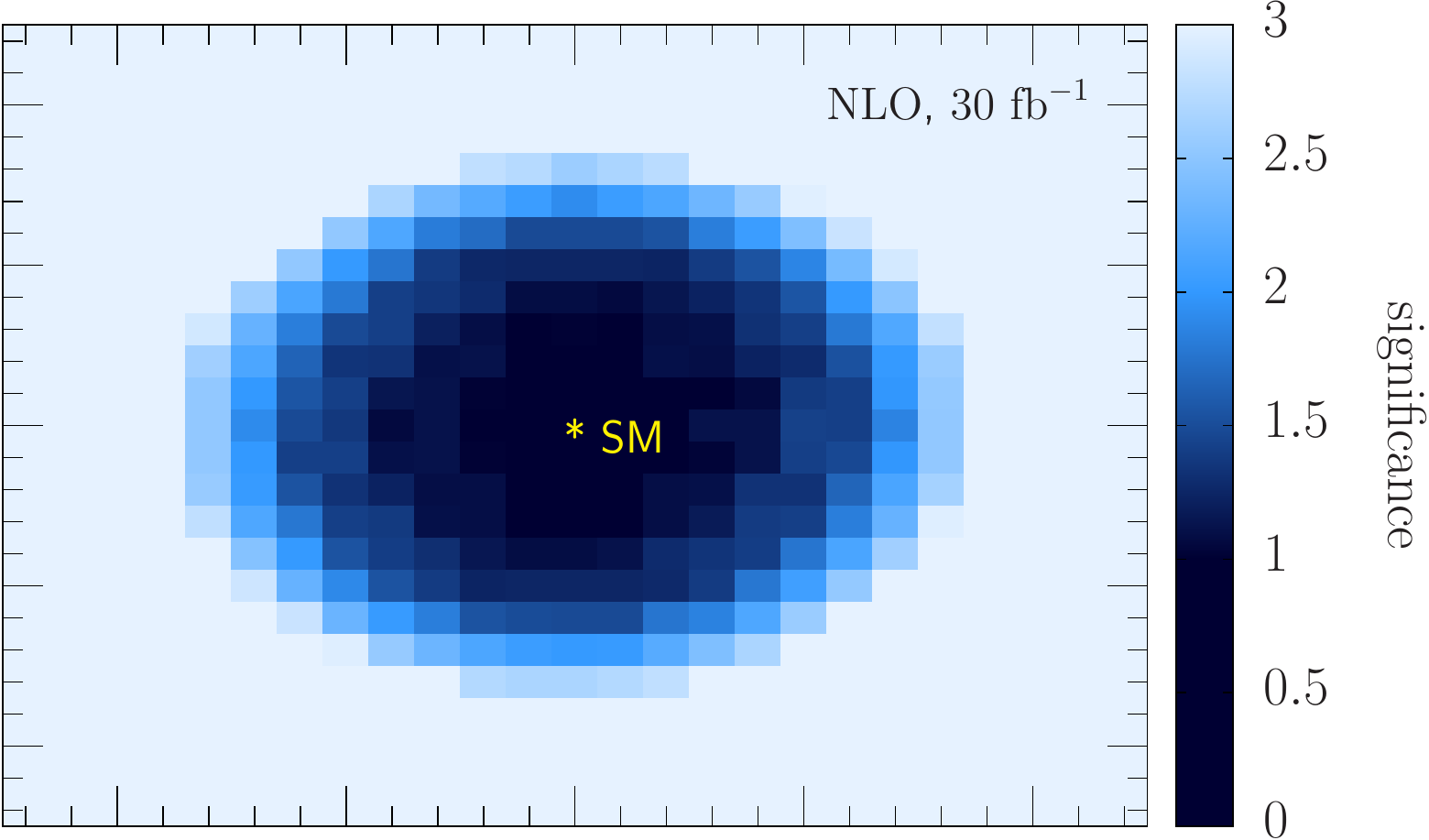}  \\
\includegraphics[scale=0.5]{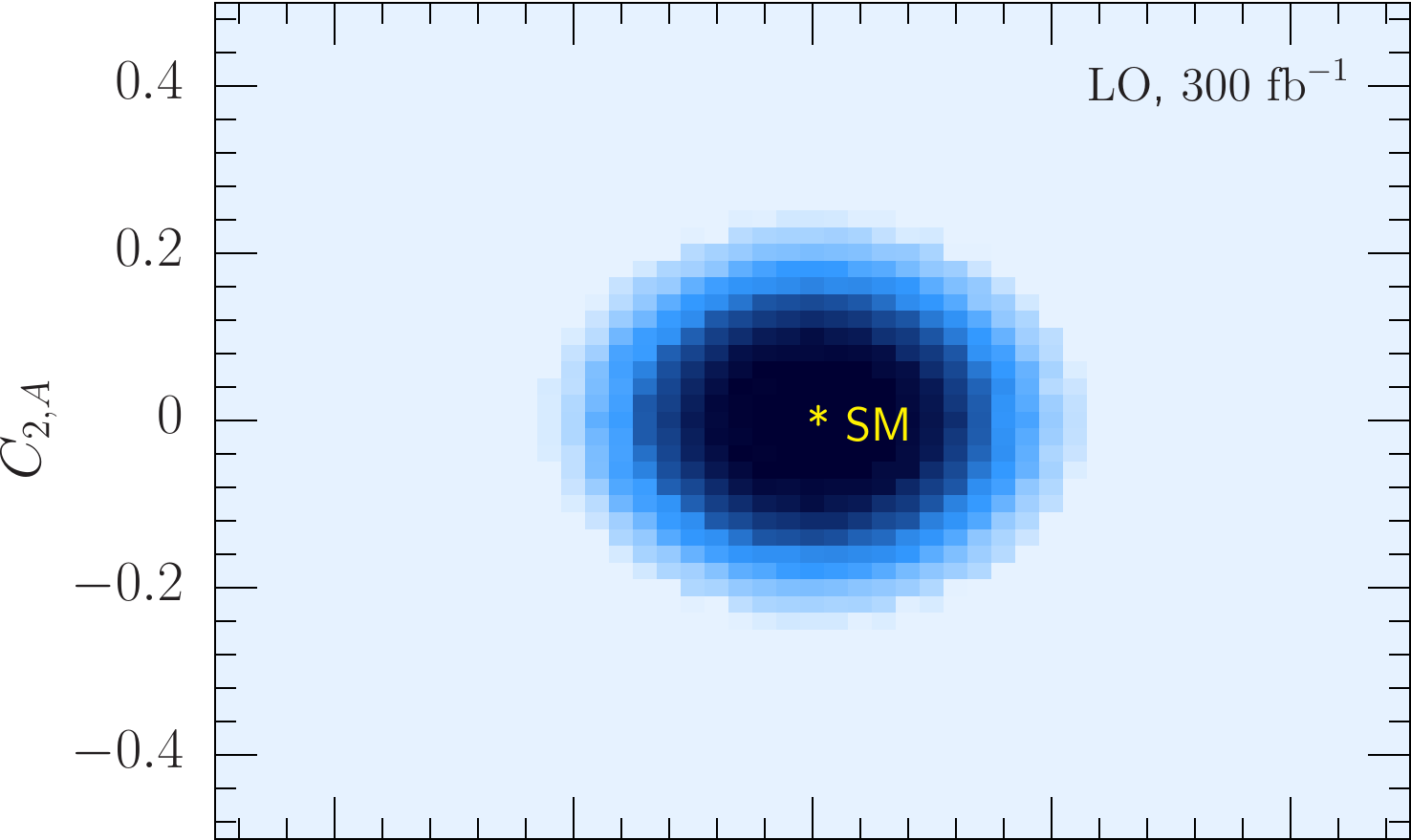} \hspace{0.1cm}
\includegraphics[scale=0.5]{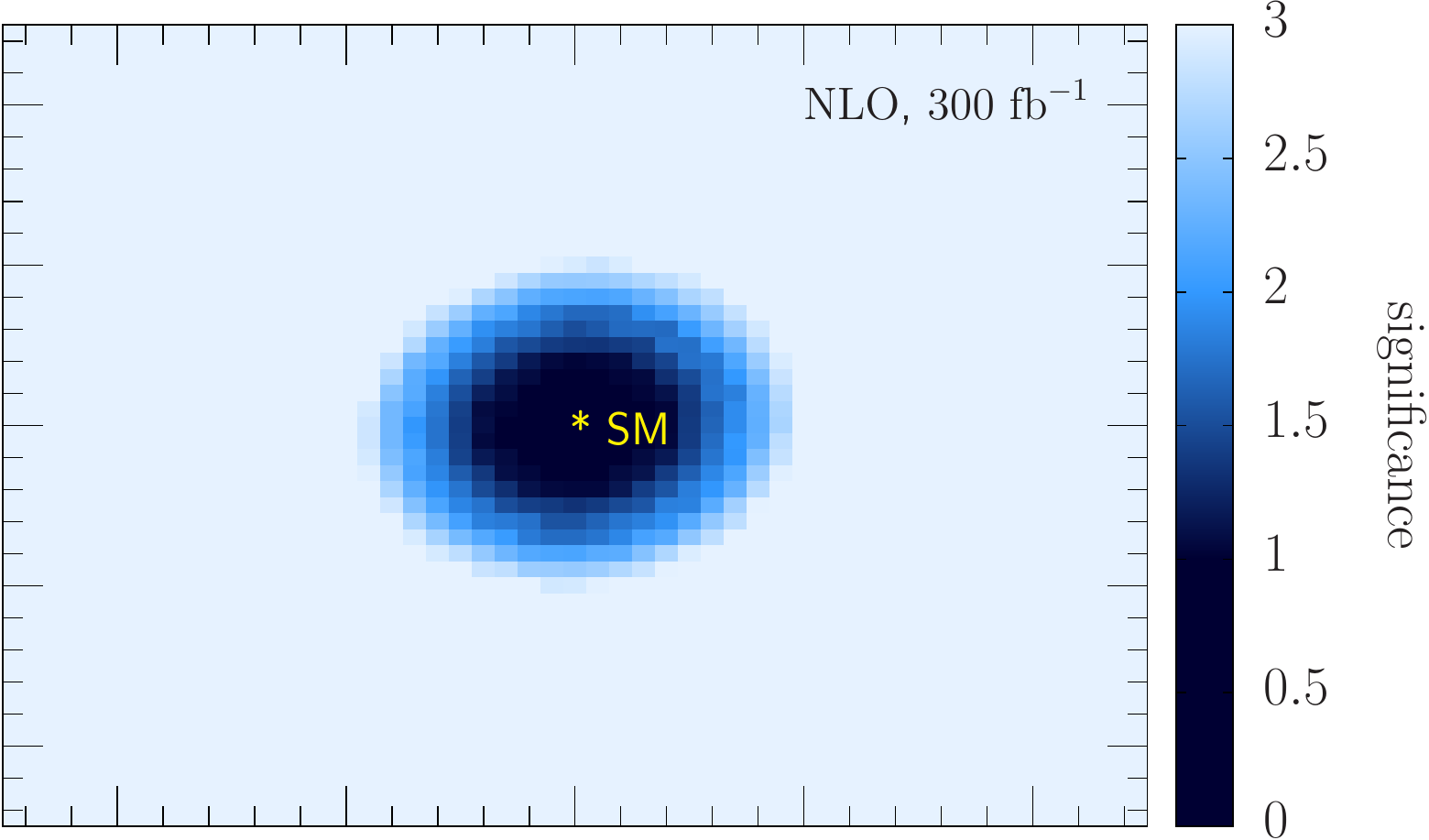}  \\
\includegraphics[scale=0.5]{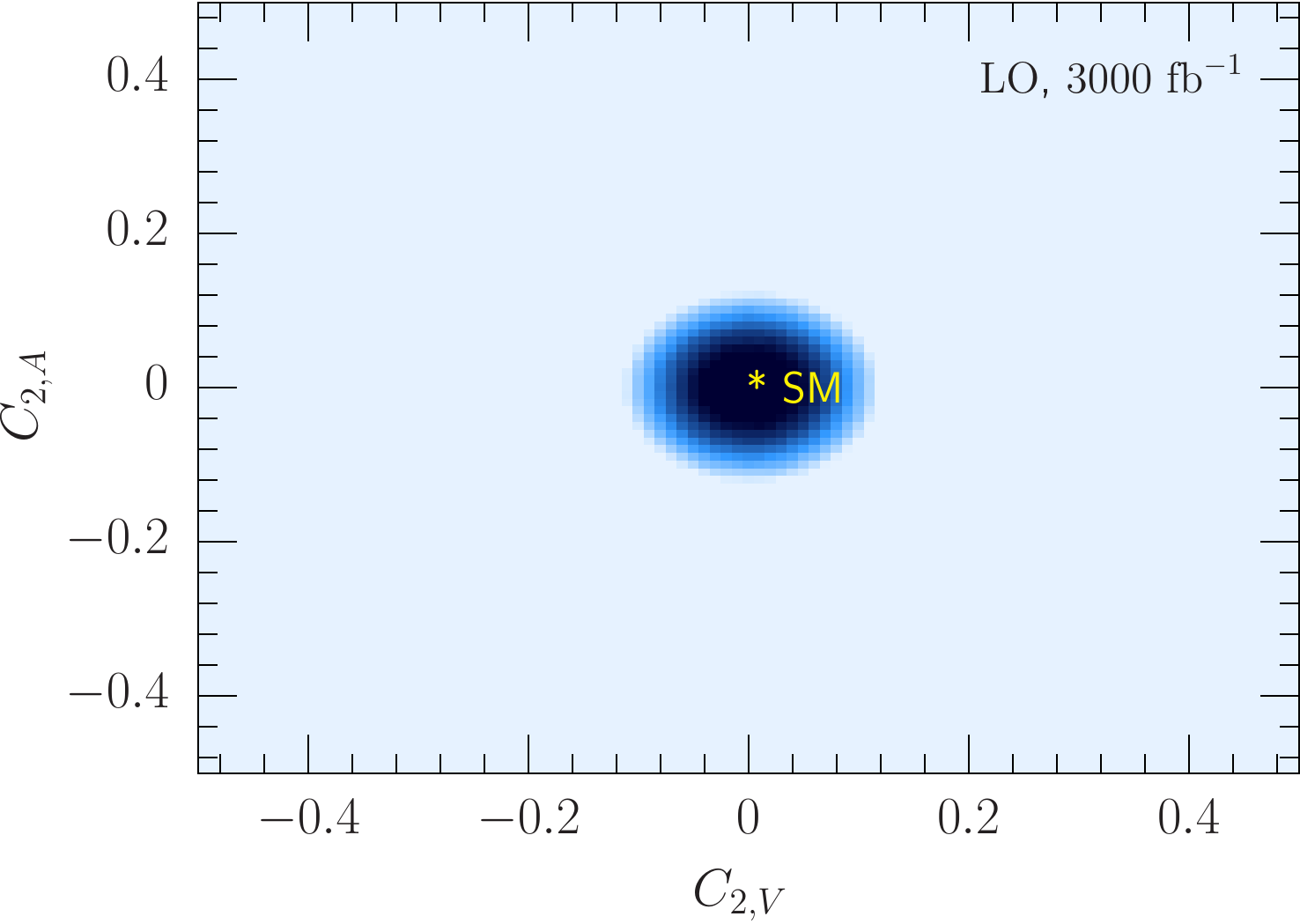} \hspace{0.1cm}
\includegraphics[scale=0.496]{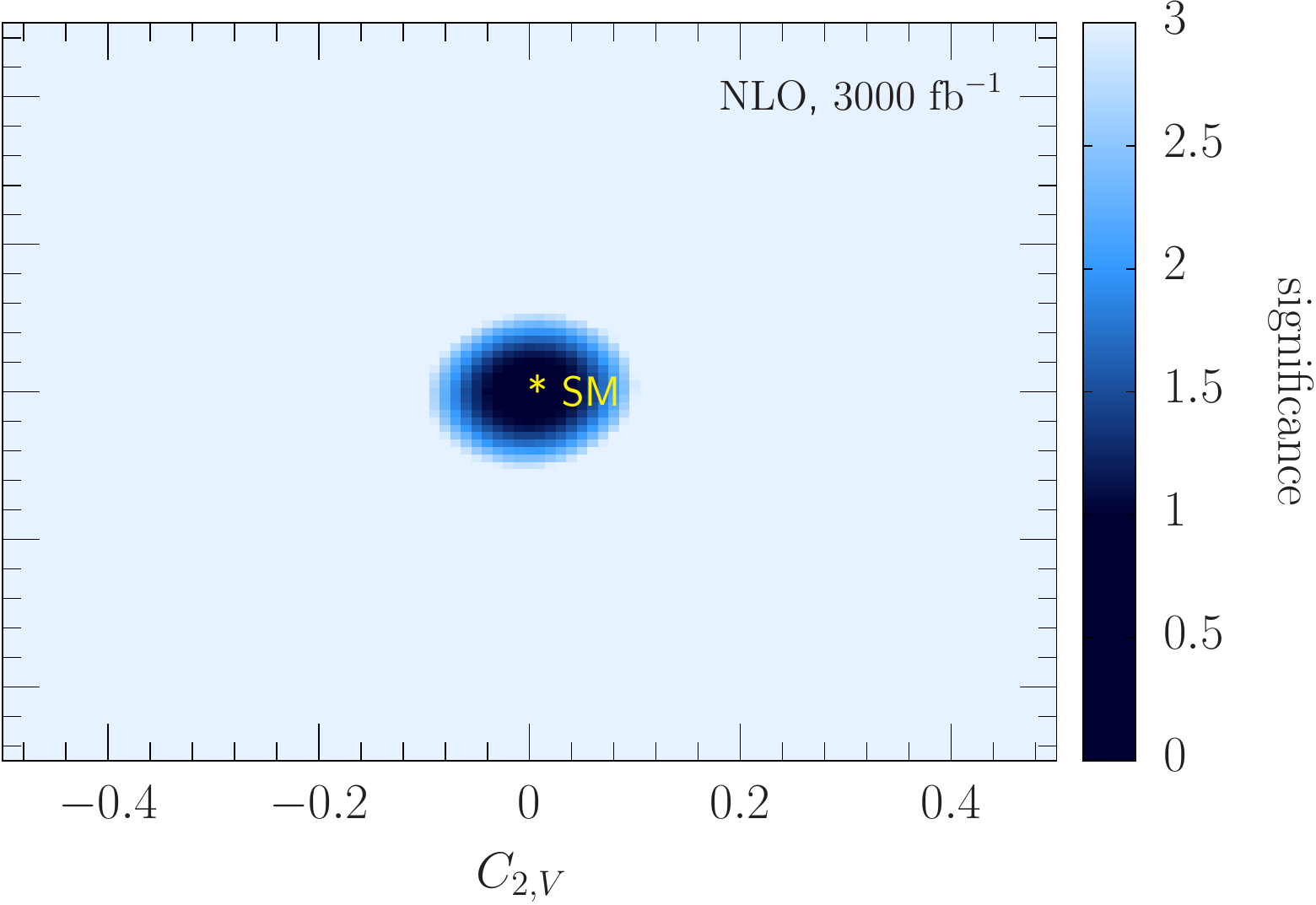}  
\caption{Significance of deviations from the SM for the dipole couplings $\CtwoV$ and $\CtwoA$ with 30, 300, and 3000~$\invfb$ of data at 
the $13$~TeV LHC. The results are obtained from rate and  $\pTZ$ shape information using LO predictions shown on the left 
and NLO QCD predictions shown on the right.\label{fig:LHC_C2}}
\end{figure}
Turning now to the dipole moment couplings $\CtwoV$ and $\CtwoA$, we focus on the constraints obtained from the $\pTZ$ distribution in figure~\ref{fig:LHC_C2}.
We choose a different color scheme to that of figure~\ref{fig:LHC_C1} to emphasize the different anomalous couplings.
We note that our LO constraints are consistent with those of Baur et. al.~\cite{Baur:2004uw}, and including NLO corrections improves the constraints by 20\%--40\%.
At this accuracy, the LHC experiments will be able to probe values of $|\CtwoVA| \lesssim 0.3$ with 30~$\invfb$ of data, improving to $|\CtwoVA| \lesssim 0.15$ with the full 300~$\invfb$ dataset 
and reaching $|\CtwoVA| \lesssim 0.08$ with the high-luminosity upgrade to 3000~$\invfb$ of data.
The effect of the NLO corrections is slightly smaller than that observed for the constraints on $\ConeVA$. 
This can be understood from our earlier observation (figure~\ref{fig:LHCdist}) that the shapes of the distributions have greater sensitivity to dipole moment couplings.
Although the overall rate is sensitive to variations in the couplings, the shape information dominates and provides the main contribution to our constraints.
Our treatment of the theoretical uncertainty involves rescaling the overall rate, so that reducing the scale uncertainty has the effect of substantially improving the contribution to the constraints coming from the rate alone, while having less of an effect on the constraints that are obtained from the shape.
Since our constraints are dominated by the latter, the improvement in the constraints when the scale uncertainty is decreased is mild.
This also explains why we see a greater improvement at lower luminosities, as this is where the rate alone is more important in obtaining the constraints.

Aside from a high-luminosity LHC and an ILC, the particle physics community is also considering the construction of a Future Circular Collider (FCC)
with proton-proton collisions at a center-of-mass energy of 100~TeV~\cite{Barletta:1956771,Dominguez,CERNwebpage}.
While this machine is usually thought of in the context of exploring a new energy regime, 
SM particles will be produced in great abundance and it is interesting to know how analyses compare to those from the clean ILC environment.
For our purposes, the $pp\to \ttbZ$ cross section at the FCC is about 50 times larger than at the 13 TeV LHC, resulting in over $10^8$ $\ttbZ$ events expected with $10~\mathrm{ab}^{-1}$ of data.
In a rough estimate we repeat our analysis for the FCC at leading-order, assuming that the theoretical uncertainty can be reduced to $\Deltaunc= 5\%$.
Our findings indicate that FCC constraints improve by factors of $3-10$ compared to the $3~\mathrm{ab}^{-1}$ LHC.
In relation to our ILC results (presented in the following section) the FCC constraints are stronger by factors of a few.

\section{ILC Constraints} \label{sec:ILC}
We now turn our attention to a future $e^+e^-$ collider.
Such a machine offers several advantages over a hadron collider, especially in performing SM precision measurements~\cite{Baer:2013cma}.
It provides an experimentally clean environment without hadronic activity in the initial state,
the collision energy is accurately known, and the possibility of polarized beams offers additional handles on analyses.
At present, the most advanced proposal for an $e^+e^-$ collider is the International Linear Collider (ILC)~\cite{Behnke:2013xla}
with a design center-of-mass energy of $\sqrt{s}=500$~GeV and the possibility for an upgrade to 1~TeV.
Precision top quark studies form an integral part of the physics program for this machine since 
at $\sqrt{s}=500$~GeV top pairs are produced numerously well above threshold. 
Moreover, the production proceeds through an intermediate $Z / \gamma$ exchange
which makes this process sensitive to electroweak top quark couplings already at leading order.

The ability of an ILC experiment to measure these couplings was analyzed in Refs.~\cite{Devetak:2010na,Amjad:2013tlv},
using as discriminating observables the cross section, the scattering angle, the forward-backward asymmetry, and the slope of the helicity angle asymmetry.
These works also take advantage of two beam polarizations to disentangle the $\ttbga$ and $\ttbZ$ couplings, and find sensitivities for the top quark couplings 
between 2--6~permille (depending on the  coupling) assuming $500~\invfb$ of data.
In this paper we use a slightly different approach to estimate our constraints.
Adopting the same collider parameters of $\sqrt{s}=500$~GeV and $500~\invfb$ of data,
we assume a simplified experimental procedure with unpolarized beams and only one single discriminating observable.
On the other hand, we account for higher-order QCD corrections and include the corresponding theoretical uncertainties in our constraints.
Hence, this procedure matches the one applied in our LHC analysis in section~\ref{sec:LHC}, enabling a direct comparison.
Following Ref.~\cite{Devetak:2010na} we use the scattering angle of the top quark to probe anomalous couplings.
We verified that this observable has a strong sensitivity to the $\ttbZ$ couplings and we consider the 
lepton+jet final state which allows a full reconstruction of the top quark momenta. We apply the following cuts to the $\ttb$  final state
\begin{eqnarray}  \label{ILCselectioncuts}
  &\pT^{\ell} > 20~\GeV, \quad  &|y^{\ell}|  < 2.5, \nonumber\\
  &\pT^{j}    > 20~\GeV, \quad &|y^{j}|  < 2.5,  \\
  &\pT^{\mathrm{miss}}> 20~\GeV, \quad  & R_j({k_\perp^{-1}})=0.4 \nonumber
\end{eqnarray}
and find $\sigma^\mathrm{NLO} = 90.0 \pm 0.8$~fb. This promises 45000 top quark pairs from $500~\invfb$ of data. 
Our prediction includes NLO QCD corrections in production and decay of the top quarks maintaining full spin correlations within the narrow-width approximation.
As pointed out in section~\ref{sec:coupl}, the anomalous dipole couplings in  $e^+e^- \to t\bar{t}$ are UV-renormalized in exactly the same way as in hadronic collisions,
given by Eq.~(\ref{C2counterterm}).
The QCD NLO corrections to top-pair production in $e^+e^-$ annihilation have been known for a long time~\cite{Jersak:1981sp,Beenakker:1991ca,Schmidt:1995mr,Kodaira:1998gt,Liu:1999fh}.
NNLO QCD corrections to the top-pair production at threshold have received significant attention in the past (see~\cite{Hoang:2000yr} and reference therein), 
and progress is being made on the $\mathrm{N^3LO}$ corrections~\cite{Beneke:2013jia,Marquard:2014pea,Beneke:2014qea} at threshold.
In the continuum, certain differential results at NNLO QCD have recently been presented recently~\cite{Gao:2014eea,Dekkers:2014hna,Gao:2014nva}.
Both the NLO scale uncertainty and the NNLO corrections at $\sqrt{s}=500$~GeV are at the level of 1\%-2\%, indicating a very good perturbative convergence~\cite{Gao:2014eea}.
NLO electroweak corrections to the overall cross section are at the level of a few percent, but Refs.~\cite{Fleischer:2003kk,Hahn:2003ab} report  
shape changes of up to 10\% in the $\cos\theta$ differential distribution.

\begin{figure}[t]
    \includegraphics[scale=0.45]{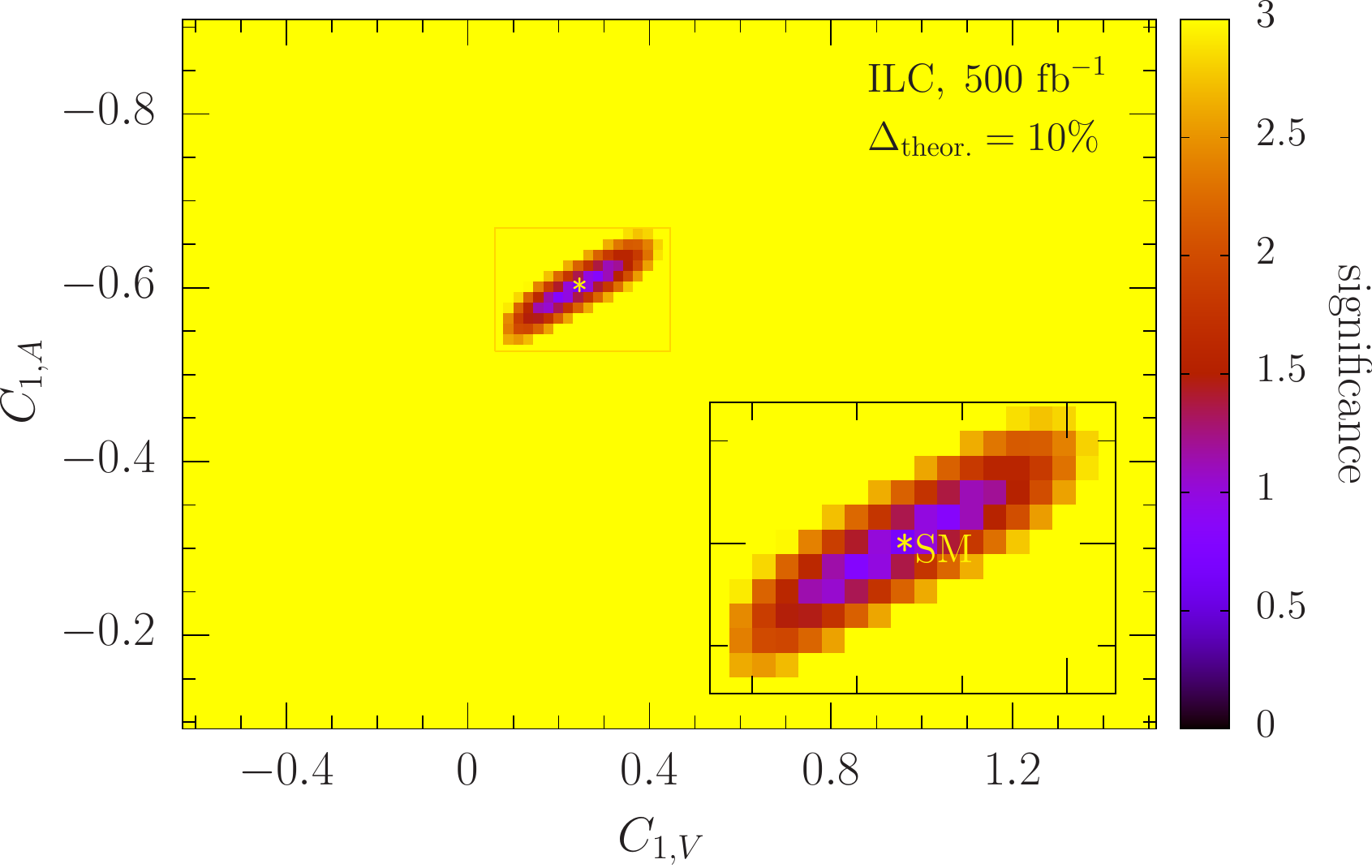} \hspace{0.2cm}
    \includegraphics[height=0.324\linewidth,width=0.5\linewidth]{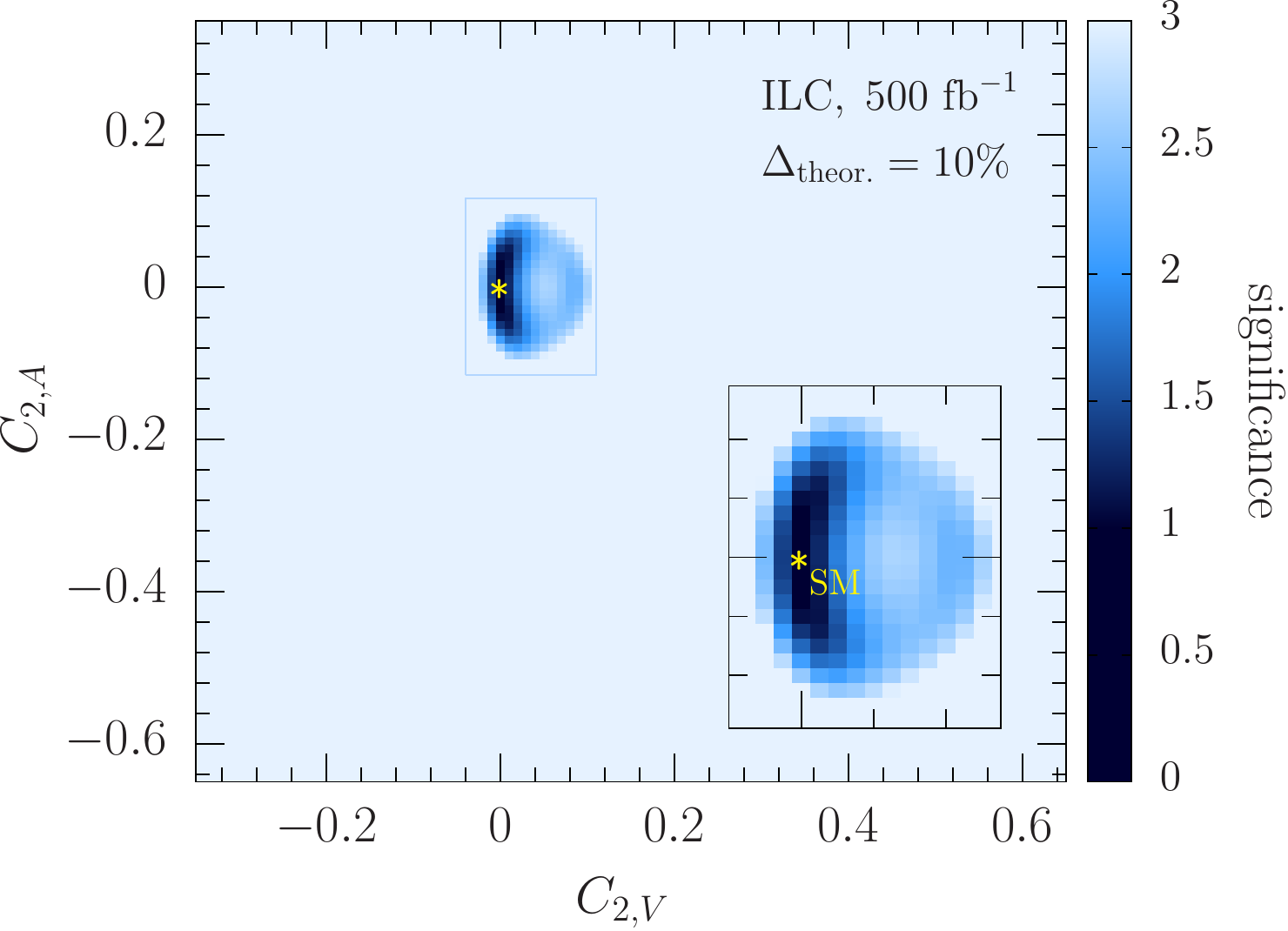}
    \caption{Sensitivities of deviations from the SM for the couplings $\ConeV$ and $\ConeA$ (left), and $\CtwoV$ and $\CtwoA$ (right)
    for the $500$~GeV ILC with $500~\invfb$ of data. 
The constraints assume NLO QCD predictions and a total uncertainty of 10\%.
For ease of comparison the scale of the large panes is kept the same as in our LHC analysis.\label{fig:epem10}
}
\end{figure}
\begin{figure}[t]
\includegraphics[scale=0.45]{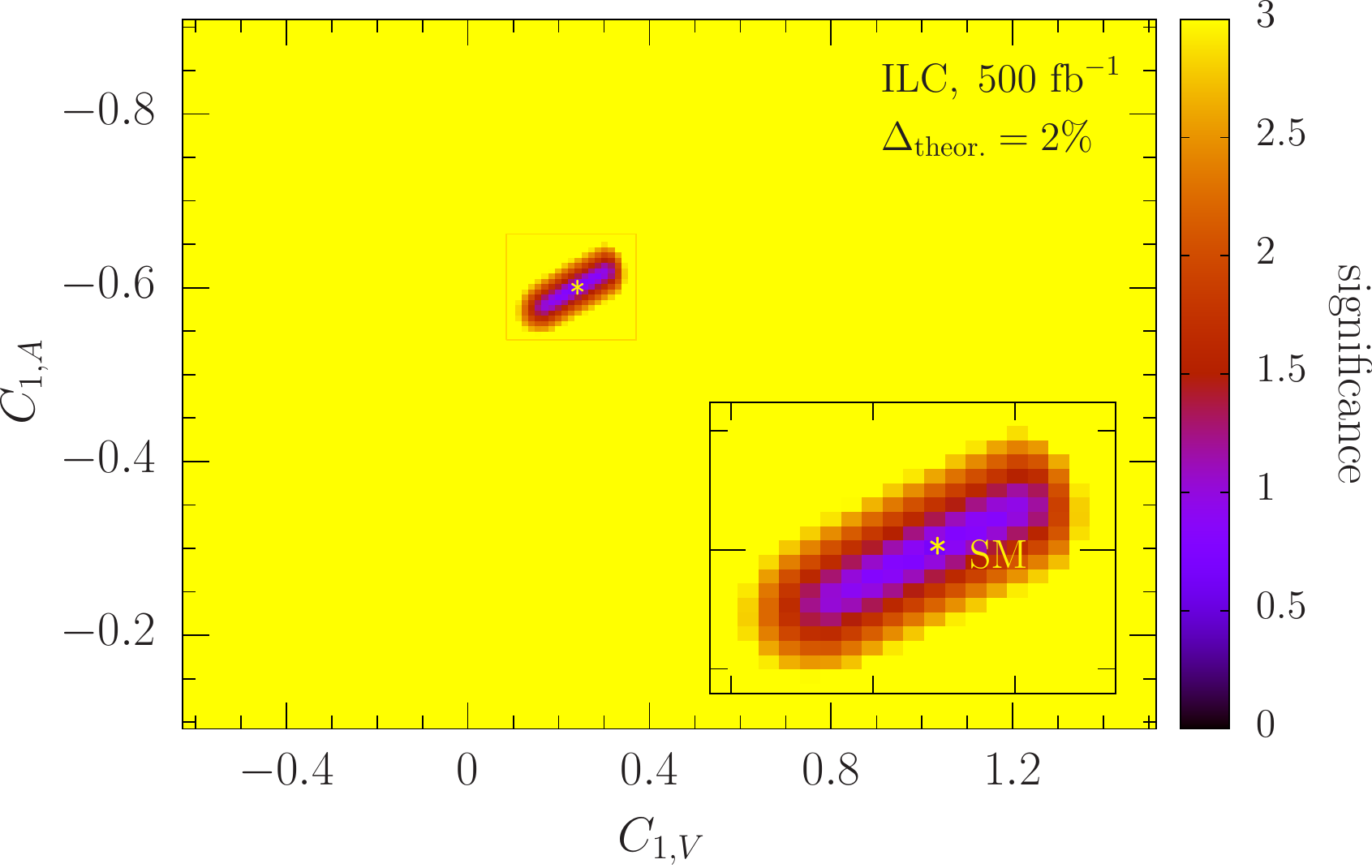} \hspace{0.2cm}
\includegraphics[height=0.324\linewidth,width=0.5\linewidth]{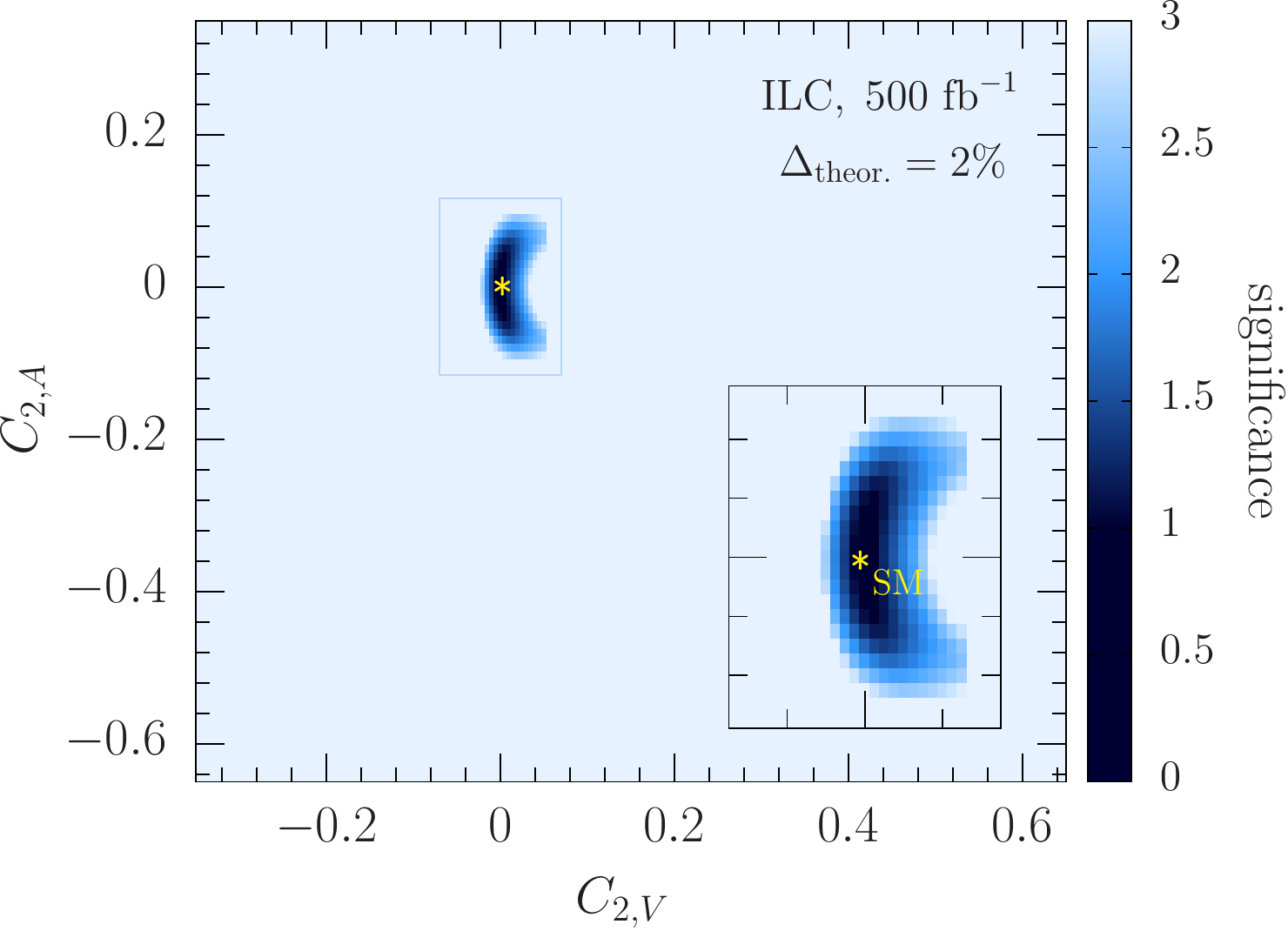}
\caption{As for figure~\protect\ref{fig:epem10}, but assuming a theoretical uncertainty of 2\%.
For ease of comparison the scale of the large panes is kept the same as in our LHC analysis.\label{fig:epem1}}
\end{figure}

    
An important part of our analysis is the inclusion of theoretical uncertainties in the constraints.
Since we neglect known electroweak corrections we will present two scenarios: 
(1) We assume a theoretical uncertainty of as much as 10\% in our log-likelihood ratio test to account for the missing electroweak corrections in our predictions;
(2) we estimate future results with 2\% uncertainty assuming that NNLO QCD and $\mathcal{O}(\alpha)$ corrections including anomalous couplings are known.
Figure~\ref{fig:epem10} presents the results for the first scenario.
Sensitivity to deviations of $\ConeV$ and $\ConeA$ is shown on the left, and the sensitivity to deviations of $\CtwoV$ and $\CtwoA$ is shown on the right.
For ease of comparison the scale of the plots is kept the same as for the LHC (figure~\ref{fig:LHC_C1},\ref{fig:LHC_C2}).
Exactly the same is shown in figure~\ref{fig:epem1} for scenario (2) with 2\% theoretical uncertainty.
In view of future results we choose to discuss the latter.
We find that vector and axial couplings can be constrained to approximately $\ConeV=0.24^{+0.08}_{-0.11}$ and $\ConeA=-0.60^{+0.04}_{-0.04}$
at the 95\% C.L.
Comparing these with the constraints presented in section~\ref{sec:LHC} for the high luminosity LHC, we see an improvement by a factor of 
three on the $\ConeA$ constraints, and an improvement by a factor of approximately six on $\ConeV$.
At both the LHC and the ILC, we see a greater sensitivity to $\ConeA$ than to $\ConeV$.
We also note that fixing one of the two couplings to its SM value improves the ILC constraints on the other coupling by a factor of about three.

Turning now to the weak dipole moments (figure~\ref{fig:epem1}, right), we find the constraints 
$|\CtwoA| \lesssim 0.08$, and $-0.02 \lesssim \CtwoV \lesssim 0.04$ at the 95\% C.L.
Comparing to the LHC constraints presented in section~\ref{sec:LHC}, we see that the constraints on $\CtwoA$ from the ILC are 
comparable to those from the LHC with $3000~\invfb$ of data, as is the upper limit on $\CtwoV$.
However, the asymmetric pattern of the ILC constraints on $\CtwoV$ allows an improvement on the lower limit of this coupling by a factor 
of three compared to the LHC. 
We emphasize that the constraints presented here are purely a statistical comparison between different 
calculations, and do not take into account experimental effects, which in general will weaken them.
Since we expect the experimental effects to be larger at the LHC than the ILC, the above discussion should be viewed as a conservative estimate 
of the relative improvement of the ILC constraints over the LHC constraints.

It is worthwhile to compare the results obtained here with those of Amjad et. al.~\cite{Amjad:2013tlv}, who also consider the ILC running at $\sqrt{s}=500$ GeV 
and with an integrated luminosity of $500~\invfb$.
Varying one coupling at a time, they find constraints at the 68\% C.L. of 4 per mille for $\ConeV$, 6 per mille for $\ConeA$, and about 1 per mille for $\CtwoV$.
They do not constrain the $\CtwoA$ coupling.
We also note that, although Ref.~\cite{Amjad:2013tlv} contains a thorough exploration of the experimental aspects, 
the quoted sensitivities do not include systematic errors.
If we assume SM values for all but one coupling, and use $\Deltaunc=2\%$, then at the same C.L. we find similar constraints on $\ConeA$.
However, our constraints are a factor of five weaker for $\ConeV$ and $\CtwoV$.
This can be understood by several differences in the analyses: 
In this paper we only use one single differential observable and assume unpolarized beams, and, 
furthermore, we include theoretical uncertainties from scale dependence in our constraints.
We stress that this last factor is an unavoidable source of systematic uncertainty which should be taken into account when projecting future constraints.
Nevertheless it is clear that a strong improvement can be obtained once polarized beams and a multi-variate analysis are employed.

\section{Conclusions} \label{sec:concl}
In this paper we have continued our work on directly constraining $\ttbZ$ couplings in collider experiments.
We discuss the study of weak magnetic and electric dipole moments, two couplings which are
an excellent probe of new physics given their small SM values.
In our analysis we consider anomalous production of top quark pairs in association with a $Z$ boson at NLO QCD and generate various differential distributions over large grids in the anomalous coupling space.
We use the obtained shape and normalization information in a log-likelihood ratio test which includes leading theoretical uncertainties
in order to reliably estimate future constraints.

We find that experiments at the 13 TeV LHC will soon be able to explore weak dipole couplings in $\ttbZ$ production,
and that including higher-order predictions boosts the range of exclusion limits.
For example, the magnetic and electric $\ttbZ$ dipole moments can be constrained to 
$\pm 0.15$ with 300~$\invfb$ in the tri-lepton final state. 
This is an improvement by 25\% compared to an analysis at leading order. 
A further improvement to $\pm 0.08$ is possible once 3000~$\invfb$ of data is accumulated.
Similar constraints can be extracted from LEP electroweak precision data, and
possibly from future $B$-physics analyses.
However these indirect limits are unavoidably diluted by assumptions on other anomalous contributions which enter the observables.
On-shell production of $\ttbZ$ at the LHC, on the other hand, is a much more direct probe and will soon yield
unambiguous and stringent results.

We also estimate constraints from an NLO QCD calculation of $\ttb$ production at the ILC, 
and find constraints of $-0.64 \lesssim \ConeA \lesssim -0.56$ and $0.13 \lesssim \ConeV \lesssim 0.32$, 
approximately a factor of three to five better than those obtainable by the LHC. 
For the weak dipole couplings we find
$-0.08 \lesssim \CtwoA \lesssim 0.08$ and $-0.02 \lesssim \CtwoV \lesssim 0.04$, which represents a factor of three improvement for $\CtwoV$.

Finally, we note that if anomalous $\ttbZ$ couplings are expressed through dimension-six operators of an effective theory they share operators with $\ttbga$ interactions.
Since the process $pp \to \ttbga$ is accessible at the LHC this offers the unique opportunity to further constrain these operators and pin down 
top quark electroweak interactions. 
Similarly, ILC experiments will be able to disentangle $Z$ and $\gamma$ interactions to top quarks if polarized beams are available.

\acknowledgments
R.R. is grateful to the CERN Theory Group for their hospitality and support during the preparation of this paper. 
Fermilab is operated by Fermi Research Alliance, LLC under Contract No. De-AC02-07CH11359 with the United States Department of Energy. 
This research used resources of the National Energy Research Scientific Computing Center (NERSC), a DOE Office of Science User Facility 
supported by the Office of Science of the U.S. Department of Energy under Contract No. DE-AC02-05CH11231.

\bibliographystyle{JHEP}
\bibliography{ttbZLHCILC}

\providecommand{\href}[2]{#2}\begingroup\raggedright\begin{thebibliography}{10}

\bibitem{ATLAS:2011nka}
{\bf ATLAS} Collaboration, {\it {Measurement of the inclusive t tbar gamma
  cross section with the ATLAS detector}},  Tech. Rep. ATLAS-CONF-2011-153,
  ATLAS-COM-CONF-2011-186, 2011.
%%CITATION = ATLAS-CONF-2011-153 ETC.;%%

\bibitem{CMS:2014wma}
{\bf CMS} Collaboration, {\it {Measurement of the inclusive top-quark pair +
  photon production cross section in the muon + jets channel in pp collisions
  at 8 TeV}},  Tech. Rep. CMS-PAS-TOP-13-011, 2014.
%%CITATION = CMS-PAS-TOP-13-011 ETC.;%%

\bibitem{Khachatryan:2014ewa}
{\bf CMS} Collaboration, V.~Khachatryan {\em et.~al.}, {\it {Measurement of top
  quark-antiquark pair production in association with a $W$ or $Z$ boson in
  $pp$ collisions at $\sqrt{s} = 8$ TeV}},
  \href{http://arXiv.org/abs/1406.7830}{{\tt 1406.7830}}.
%%CITATION = ARXIV:1406.7830;%%

\bibitem{ATLAS-CONF-2014-038}
{\bf ATLAS} Collaboration, {\it {Evidence for the associated production of a
  vector boson (W, Z) and top quark pair in the dilepton and trilepton channels
  in pp collision data at $\sqrt{s}=8$ TeV collected by the ATLAS detector at
  the LHC}},  Tech. Rep. ATLAS-CONF-2014-038, 2014.

\bibitem{altarelli:1990zd}
G.~Altarelli and R.~Barbieri, {\it {Vacuum polarization effects of new physics
  on electroweak processes}},  {\em Phys.Lett.} {\bf B253} (1991) 161--167.
%%CITATION = PHLTA,B253,161;%%

\bibitem{Altarelli:1991fk}
G.~Altarelli, R.~Barbieri and S.~Jadach, {\it {Toward a model independent
  analysis of electroweak data}},  {\em Nucl.Phys.} {\bf B369} (1992) 3--32.
%%CITATION = NUPHA,B369,3;%%

\bibitem{Altarelli:1993sz}
G.~Altarelli, R.~Barbieri and F.~Caravaglios, {\it {Nonstandard analysis of
  electroweak precision data}},  {\em Nucl.Phys.} {\bf B405} (1993) 3--23.
%%CITATION = NUPHA,B405,3;%%

\bibitem{Eboli:1997kd}
O.~J. Eboli, M.~Gonzalez-Garcia and S.~Novaes, {\it {Limits on anomalous top
  couplings from Z pole physics}},  {\em Phys.Lett.} {\bf B415} (1997) 75--82
  [\href{http://arXiv.org/abs/hep-ph/9704400}{{\tt hep-ph/9704400}}].
%%CITATION = HEP-PH/9704400;%%

\bibitem{Brod:2014hsa}
J.~Brod, A.~Greljo, E.~Stamou and P.~Uttayarat, {\it {Probing anomalous $t\bar
  t Z$ interactions with rare meson decays}},
  \href{http://arXiv.org/abs/1408.0792}{{\tt 1408.0792}}.
%%CITATION = ARXIV:1408.0792;%%

\bibitem{Kamenik:2011dk}
J.~F. Kamenik, M.~Papucci and A.~Weiler, {\it {Constraining the dipole moments
  of the top quark}},  {\em Phys.Rev.} {\bf D85} (2012) 071501
  [\href{http://arXiv.org/abs/1107.3143}{{\tt 1107.3143}}].
%%CITATION = ARXIV:1107.3143;%%

\bibitem{Bouzas:2012av}
A.~O. Bouzas and F.~Larios, {\it {Electromagnetic dipole moments of the Top
  quark}},  {\em Phys.Rev.} {\bf D87} (2013), no.~7 074015
  [\href{http://arXiv.org/abs/1212.6575}{{\tt 1212.6575}}].
%%CITATION = ARXIV:1212.6575;%%

\bibitem{Baer:2013cma}
H.~Baer, T.~Barklow, K.~Fujii, Y.~Gao, A.~Hoang {\em et.~al.}, {\it {The
  International Linear Collider Technical Design Report - Volume 2: Physics}},
  \href{http://arXiv.org/abs/1306.6352}{{\tt 1306.6352}}.
%%CITATION = ARXIV:1306.6352;%%

\bibitem{Hollik:1998vz}
W.~Hollik, J.~I. Illana, S.~Rigolin, C.~Schappacher and D.~Stockinger, {\it
  {Top dipole form-factors and loop induced CP violation in supersymmetry}},
  {\em Nucl.Phys.} {\bf B551} (1999) 3--40
  [\href{http://arXiv.org/abs/hep-ph/9812298}{{\tt hep-ph/9812298}}].
%%CITATION = HEP-PH/9812298;%%

\bibitem{Agashe:2006wa}
K.~Agashe, G.~Perez and A.~Soni, {\it {Collider Signals of Top Quark Flavor
  Violation from a Warped Extra Dimension}},  {\em Phys.Rev.} {\bf D75} (2007)
  015002 [\href{http://arXiv.org/abs/hep-ph/0606293}{{\tt hep-ph/0606293}}].
%%CITATION = HEP-PH/0606293;%%

\bibitem{Kagan:2009bn}
A.~L. Kagan, G.~Perez, T.~Volansky and J.~Zupan, {\it {General Minimal Flavor
  Violation}},  {\em Phys.Rev.} {\bf D80} (2009) 076002
  [\href{http://arXiv.org/abs/0903.1794}{{\tt 0903.1794}}].
%%CITATION = ARXIV:0903.1794;%%

\bibitem{PhysRevD.82.055001}
T.~Ibrahim and P.~Nath, {\it Top quark electric dipole moment in a minimal
  supersymmetric standard model extension with vectorlike multiplets},  {\em
  Phys. Rev. D} {\bf 82} (Sep, 2010) 055001.

\bibitem{PhysRevD.84.015003}
T.~Ibrahim and P.~Nath, {\it Chromoelectric dipole moment of the top quark in
  models with vectorlike multiplets},  {\em Phys. Rev. D} {\bf 84} (Jul, 2011)
  015003.

\bibitem{Grojean:2013qca}
C.~Grojean, O.~Matsedonskyi and G.~Panico, {\it {Light top partners and
  precision physics}},  {\em JHEP} {\bf 1310} (2013) 160
  [\href{http://arXiv.org/abs/1306.4655}{{\tt 1306.4655}}].
%%CITATION = ARXIV:1306.4655;%%

\bibitem{Richard:2013pwa}
F.~Richard, {\it {Can LHC observe an anomaly in $t\bar tZ$ production?}},
  \href{http://arXiv.org/abs/1304.3594}{{\tt 1304.3594}}.
%%CITATION = ARXIV:1304.3594;%%

\bibitem{Baur:2004uw}
U.~Baur, A.~Juste, L.~Orr and D.~Rainwater, {\it {Probing electroweak top quark
  couplings at hadron colliders}},  {\em Phys.Rev.} {\bf D71} (2005) 054013
  [\href{http://arXiv.org/abs/hep-ph/0412021}{{\tt hep-ph/0412021}}].
%%CITATION = HEP-PH/0412021;%%

\bibitem{Baur:2005wi}
U.~Baur, A.~Juste, D.~Rainwater and L.~Orr, {\it {Improved measurement of $ttZ$
  couplings at the CERN LHC}},  {\em Phys.Rev.} {\bf D73} (2006) 034016
  [\href{http://arXiv.org/abs/hep-ph/0512262}{{\tt hep-ph/0512262}}].
%%CITATION = HEP-PH/0512262;%%

\bibitem{Devetak:2010na}
E.~Devetak, A.~Nomerotski and M.~Peskin, {\it {Top quark anomalous couplings at
  the International Linear Collider}},  {\em Phys.Rev.} {\bf D84} (2011) 034029
  [\href{http://arXiv.org/abs/1005.1756}{{\tt 1005.1756}}].
%%CITATION = ARXIV:1005.1756;%%

\bibitem{Amjad:2013tlv}
M.~Amjad, M.~Boronat, T.~Frisson, I.~Garcia, R.~Poschl {\em et.~al.}, {\it {A
  precise determination of top quark electro-weak couplings at the ILC
  operating at $\sqrt{s}=500$ GeV}},
  \href{http://arXiv.org/abs/1307.8102}{{\tt 1307.8102}}.
%%CITATION = ARXIV:1307.8102;%%

\bibitem{Rontsch:2014cca}
R.~Rontsch and M.~Schulze, {\it {Constraining couplings of top quarks to the Z
  boson in $ t\overline{t} $ + Z production at the LHC}},  {\em JHEP} {\bf
  1407} (2014) 091 [\href{http://arXiv.org/abs/1404.1005}{{\tt 1404.1005}}].
%%CITATION = ARXIV:1404.1005;%%

\bibitem{Bernabeu:1995gs}
J.~Bernabeu, D.~Comelli, L.~Lavoura and J.~P. Silva, {\it {Weak magnetic dipole
  moments in two Higgs doublet models}},  {\em Phys.Rev.} {\bf D53} (1996)
  5222--5232 [\href{http://arXiv.org/abs/hep-ph/9509416}{{\tt
  hep-ph/9509416}}].
%%CITATION = HEP-PH/9509416;%%

\bibitem{Czarnecki:1996rx}
A.~Czarnecki and B.~Krause, {\it {On the dipole moments of fermions at two
  loops}},  {\em Acta Phys.Polon.} {\bf B28} (1997) 829--834
  [\href{http://arXiv.org/abs/hep-ph/9611299}{{\tt hep-ph/9611299}}].
%%CITATION = HEP-PH/9611299;%%

\bibitem{AguilarSaavedra:2008zc}
J.~Aguilar-Saavedra, {\it {A Minimal set of top anomalous couplings}},  {\em
  Nucl.Phys.} {\bf B812} (2009) 181--204
  [\href{http://arXiv.org/abs/0811.3842}{{\tt 0811.3842}}].
%%CITATION = ARXIV:0811.3842;%%

\bibitem{Lazopoulos:2008de}
A.~Lazopoulos, T.~McElmurry, K.~Melnikov and F.~Petriello, {\it
  {Next-to-leading order QCD corrections to $t \bar{t} Z$ production at the
  LHC}},  {\em Phys.Lett.} {\bf B666} (2008) 62--65
  [\href{http://arXiv.org/abs/0804.2220}{{\tt 0804.2220}}].
%%CITATION = ARXIV:0804.2220;%%

\bibitem{Kardos:2011na}
A.~Kardos, Z.~Trocsanyi and C.~Papadopoulos, {\it {Top quark pair production in
  association with a Z-boson at NLO accuracy}},  {\em Phys.Rev.} {\bf D85}
  (2012) 054015 [\href{http://arXiv.org/abs/1111.0610}{{\tt 1111.0610}}].
%%CITATION = ARXIV:1111.0610;%%

\bibitem{Martin:2009iq}
A.~Martin, W.~Stirling, R.~Thorne and G.~Watt, {\it {Parton distributions for
  the LHC}},  {\em Eur.Phys.J.} {\bf C63} (2009) 189--285
  [\href{http://arXiv.org/abs/0901.0002}{{\tt 0901.0002}}].
%%CITATION = ARXIV:0901.0002;%%

\bibitem{Cacciari:2008gp}
M.~Cacciari, G.~P. Salam and G.~Soyez, {\it {The Anti-k(t) jet clustering
  algorithm}},  {\em JHEP} {\bf 0804} (2008) 063
  [\href{http://arXiv.org/abs/0802.1189}{{\tt 0802.1189}}].
%%CITATION = ARXIV:0802.1189;%%

\bibitem{Barletta:1956771}
W.~Barletta, M.~Battaglia, M.~Klute, M.~Mangano, S.~Prestemon, L.~Rossi and
  P.~Skands, {\it {Future Hadron Colliders from physics perspectives to
  technology R and D}},  {\em Nucl. Instrum. Methods Phys. Res., A} {\bf 764}
  (Jan, 2014) 352--368. 17 p.

\bibitem{Dominguez}
C.~Dominguez and F.~Zimmerman, {\it {Proceedings, 4th International Particle
  Accelerator Conference (IPAC 2013)}},  {\em JACoW} (2013) 1442.
%%CITATION = INSPIRE-1241922;%%

\bibitem{CERNwebpage}
https://espace2013.cern.ch/fcc/Pages/Hadron-Collider.aspx.

\bibitem{Behnke:2013xla}
T.~Behnke, J.~E. Brau, B.~Foster, J.~Fuster, M.~Harrison {\em et.~al.}, {\it
  {The International Linear Collider Technical Design Report - Volume 1:
  Executive Summary}},  \href{http://arXiv.org/abs/1306.6327}{{\tt 1306.6327}}.
%%CITATION = ARXIV:1306.6327;%%

\bibitem{Jersak:1981sp}
J.~Jersak, E.~Laermann and P.~Zerwas, {\it {Electroweak Production of Heavy
  Quarks in e+ e- Annihilation}},  {\em Phys.Rev.} {\bf D25} (1982) 1218.
%%CITATION = PHRVA,D25,1218;%%

\bibitem{Beenakker:1991ca}
W.~Beenakker, S.~van~der Marck and W.~Hollik, {\it {e+ e- annihilation into
  heavy fermion pairs at high-energy colliders}},  {\em Nucl.Phys.} {\bf B365}
  (1991) 24--78.
%%CITATION = NUPHA,B365,24;%%

\bibitem{Schmidt:1995mr}
C.~R. Schmidt, {\it {Top quark production and decay at next-to-leading order in
  e+ e- annihilation}},  {\em Phys.Rev.} {\bf D54} (1996) 3250--3265
  [\href{http://arXiv.org/abs/hep-ph/9504434}{{\tt hep-ph/9504434}}].
%%CITATION = HEP-PH/9504434;%%

\bibitem{Kodaira:1998gt}
J.~Kodaira, T.~Nasuno and S.~J. Parke, {\it {QCD corrections to spin
  correlations in top quark production at lepton colliders}},  {\em Phys.Rev.}
  {\bf D59} (1998) 014023 [\href{http://arXiv.org/abs/hep-ph/9807209}{{\tt
  hep-ph/9807209}}].
%%CITATION = HEP-PH/9807209;%%

\bibitem{Liu:1999fh}
H.~X. Liu, C.~S. Li and Z.~J. Xiao, {\it {O(alpha($s$) ) QCD corrections to
  spin correlations in $e^{-} e^{+} \to t \bar{t}$ process at the NLC}},  {\em
  Phys.Lett.} {\bf B458} (1999) 393--401
  [\href{http://arXiv.org/abs/hep-ph/9901205}{{\tt hep-ph/9901205}}].
%%CITATION = HEP-PH/9901205;%%

\bibitem{Hoang:2000yr}
A.~Hoang, M.~Beneke, K.~Melnikov, T.~Nagano, A.~Ota {\em et.~al.}, {\it {Top -
  anti-top pair production close to threshold: Synopsis of recent NNLO
  results}},  {\em Eur.Phys.J.direct} {\bf C2} (2000) 1
  [\href{http://arXiv.org/abs/hep-ph/0001286}{{\tt hep-ph/0001286}}].
%%CITATION = HEP-PH/0001286;%%

\bibitem{Beneke:2013jia}
M.~Beneke, Y.~Kiyo and K.~Schuller, {\it {Third-order correction to top-quark
  pair production near threshold I. Effective theory set-up and matching
  coefficients}},  \href{http://arXiv.org/abs/1312.4791}{{\tt 1312.4791}}.
%%CITATION = ARXIV:1312.4791;%%

\bibitem{Marquard:2014pea}
P.~Marquard, J.~H. Piclum, D.~Seidel and M.~Steinhauser, {\it {Three-loop
  matching of the vector current}},  {\em Phys.Rev.} {\bf D89} (2014) 034027
  [\href{http://arXiv.org/abs/1401.3004}{{\tt 1401.3004}}].
%%CITATION = ARXIV:1401.3004;%%

\bibitem{Beneke:2014qea}
M.~Beneke, Y.~Kiyo, P.~Marquard, A.~Penin, J.~Piclum {\em et.~al.}, {\it
  {Leptonic decay of the Upsilon(1S) meson at third order in QCD}},  {\em
  Phys.Rev.Lett.} {\bf 112} (2014) 151801
  [\href{http://arXiv.org/abs/1401.3005}{{\tt 1401.3005}}].
%%CITATION = ARXIV:1401.3005;%%

\bibitem{Gao:2014eea}
J.~Gao and H.~X. Zhu, {\it {Top-quark forward-backward asymmetry in e+e-
  annihilation at NNLO in QCD}},  \href{http://arXiv.org/abs/1410.3165}{{\tt
  1410.3165}}.
%%CITATION = ARXIV:1410.3165;%%

\bibitem{Dekkers:2014hna}
O.~Dekkers and W.~Bernreuther, {\it {The real-virtual antenna functions for $S
  \to Q\bar{Q} X$ at NNLO QCD}},  {\em Phys.Lett.} {\bf B738} (2014) 325--333
  [\href{http://arXiv.org/abs/1409.3124}{{\tt 1409.3124}}].
%%CITATION = ARXIV:1409.3124;%%

\bibitem{Gao:2014nva}
J.~Gao and H.~X. Zhu, {\it {Electroweak prodution of top-quark pairs in e+e-
  annihilation at NNLO in QCD: the vector contributions}},
  \href{http://arXiv.org/abs/1408.5150}{{\tt 1408.5150}}.
%%CITATION = ARXIV:1408.5150;%%

\bibitem{Fleischer:2003kk}
J.~Fleischer, A.~Leike, T.~Riemann and A.~Werthenbach, {\it {Electroweak one
  loop corrections for e+ e- annihilation into t anti-top including hard
  bremsstrahlung}},  {\em Eur.Phys.J.} {\bf C31} (2003) 37--56
  [\href{http://arXiv.org/abs/hep-ph/0302259}{{\tt hep-ph/0302259}}].
%%CITATION = HEP-PH/0302259;%%

\bibitem{Hahn:2003ab}
T.~Hahn, W.~Hollik, A.~Lorca, T.~Riemann and A.~Werthenbach, {\it {O(alpha)
  electroweak corrections to the processes e+ e- $\to$ tau- tau+, c anti-c, b
  anti-b, t anti-t: A Comparison}},
  \href{http://arXiv.org/abs/hep-ph/0307132}{{\tt hep-ph/0307132}}.
%%CITATION = HEP-PH/0307132;%%

\end{thebibliography}\endgroup
\end{document}